\documentclass[journal=jpcbfk,manuscript=article]{achemso}
\setkeys{acs}{etalmode=truncate,maxauthors=0}

\usepackage{fullpage}
\usepackage{graphicx,graphics,tabularx,wrapfig,subfigure,amsmath,footmisc,verbatim,calc,epsfig} 
\usepackage[sort&compress]{natbib}
\usepackage[normalem]{ulem}
\usepackage{setspace}
\usepackage{amssymb}
\usepackage{makecell}
\usepackage{multicol}
\usepackage{relsize}
\usepackage{tikz}
\usepackage{enumitem}
\usepackage[utf8]{inputenc}	
\usepackage[version=4,arrows=pgf-filled,
textfontname=sffamily,
mathfontname=mathsf]{mhchem}
\usepackage{physics}
\usepackage{xr}
\usepackage{booktabs}

\title{The Stability of Rutile Oxides for Oxygen Evolution Catalysis: From Mechanistic Understanding to New Descriptors}

\author{Katarina Kretschmer}
\affiliation[FZJ]{Theory and Computation of Energy Materials (IET-3), Institute of Energy Technologies, Forschungszentrum Jülich GmbH, 52425 Jülich, Germany}

\author{Michael H. Eikerling}
\affiliation[FZJ]{Theory and Computation of Energy Materials (IET-3), Institute of Energy Technologies, Forschungszentrum Jülich GmbH, 52425 Jülich, Germany}
\alsoaffiliation[RWTH]{Chair of Theory and Computation of Energy Materials, Faculty of Georesources and Materials Engineering, RWTH Aachen University, 52062 Aachen, Germany}

\author{Tobias Binninger}
\affiliation[FZJ]{Theory and Computation of Energy Materials (IET-3), Institute of Energy Technologies, Forschungszentrum Jülich GmbH, 52425 Jülich, Germany}
\email{t.binninger@fz-juelich.de}

\begin{document}
\doublespacing
\begin{abstract}
\thispagestyle{empty}
Polymer electrolyte water electrolysis is considered as a pivotal technology for green hydrogen production. Iridium dioxide (IrO$_{2}$) serves as the benchmark electrocatalyst for the anodic oxygen evolution reaction (OER) in these cells. However, the scarcity and cost of iridium drives efforts to find durable and cost-effective replacements or drastically reduce the iridium loading. While ruthenium dioxide (RuO$_{2}$) offers comparable activity, its practical utility is impacted by its stability being inferior to IrO$_{2}$. The presented computational study analyzes the metal--oxygen bond strengths in bulk IrO$_{2}$ and RuO$_{2}$ to provide insights into the electronic origins of their differing stabilities. Through a comparative analysis of the electronic structure of bulk rutile IrO$_{2}$, RuO$_{2}$, and selected MO$_{2}$ systems, it is able to discriminate the covalent and ionic contributions to the metal--oxygen bond using crystal-orbital-derived electronic descriptors. 

\end{abstract}
\newpage
\setcounter{page}{1}
\section{Introduction}
Polymer electrolyte water electrolysis (PEWE) is deemed a key technology for the green hydrogen economy, offering a pathway for sustainable generation of highly-purified hydrogen \cite{smolinka2018,kiemel2021,chu2017,minke2021,kani2024,chatenet2022,buttler2018,barbir2005,veziroglu2008}. The oxygen evolution reaction (OER) at the anode of PEWE produces molecular oxygen via electrochemical oxidation of water. This multi-step reaction requires highly active electrocatalysts to overcome substantial kinetic overpotentials inherent to the OER mechanism. Besides providing the required activity for the OER, suitable electrocatalysts must exhibit sufficient stability to withstand the harsh acidic and highly oxidative environment prevailing the PEWE anode catalyst layer \cite{shi2020, schweinar2020}.\\

Iridium dioxide (IrO$_{2}$) and ruthenium dioxide (RuO$_{2}$) are known as the most promising OER electrocatalysts, exhibiting superior performance and favorable activity-stability trade-offs compared to other platinum-group oxides \cite{shi2020,schweinar2020,exner2023,mccrory2015,mccrory2013,seitz2016,jones2024,danilovic2014,weber2021}. While both are viable candidates for PEWE, IrO$_{2}$ demonstrates superior electrochemical stability under the acidic conditions and high anodic potentials prevailing in the cell \cite{danilovic2014,shi2020,schweinar2020,exner2023,mccrory2013,mccrory2015,jones2024,geiger2018,weber2021}. However, the scarcity of iridium in the upper crust of the earth ($\sim$\,0.05\,ppb) \cite{wedepohl1995} and its high cost poses a critical barrier to the deployment of PEWE \cite{clapp2023,minke2021,hubert2022,hoffmann_britannica_PGM}. This constraint necessitates alternative electrocatalysts to be developed, that provide comparable catalytic activity and stability while being economically viable \cite{katsounaros2014,stamenkovic2017,song2020}.\\

Although research into OER materials has intensified over recent years, identifying alternative anode catalysts for PEWE remains an open challenge \cite{lee2022,rossmeisl2007,rossmeisl2005,kasian2018,liang2021,grimaud2016,rong2016,evazzade2023,lu2024,maraschin2026,ping2015, danilovic2014,cherevko2014,cherevko2016,kahk2014,binninger2015,sargeant2022,geuss2024,weber2021,lee2012,sanchezcasalongue2014,stoerzinger2014,kuo2017,saveleva2018,xu2015,gauthier2017,briquet2017,exner2019,nong2020,gonzalez2021,binninger2022}.  A recent experimental investigation shows that while amorphous IrO$_\mathrm{x}$ exhibits higher activity, rutile IrO$_{2}$ is two orders of magnitude more stable under OER conditions \cite{geuss2024}. On the other hand, strategies pursuing ultra-low iridium loading often lead to increased dissolution, compromising long-term application \cite{cherevko2014,geuss2024,lim2025}.\\

An earlier theoretical study demonstrated, on the basis of fundamental thermodynamic considerations, that the presence of oxygen anions in the metal oxide lattice renders every metal oxide thermodynamically unstable, independent of pH \cite{binninger2015}. A subsequent study suggested that the OER mechanism on the IrO$_{2}$ (110) surface warrants reconsideration. Rather than following a mononuclear mechanism, it is proposed the OER on IrO$_{2}$ (110) to proceed via a binuclear mechanism (BM) involving adjacent surface oxygen atoms as adsorption sites, while the iridium atoms in the surface underneath play an indirect role allowing them to retain their saturated sixfold coordination throughout all stages of the reaction. Molecular oxygen is formed through an Ir--OOOO--Ir transition state by association of the terminal oxygen atoms of two adjacent Ir--OO surface entities, leaving two intact Ir--O entities at the surface. This contrasts with the mononuclear mechanism, in which O$_{2}$ evolves via cleavage of the Ir--O bond within an Ir--OO entity. The proposed binuclear pathway assumes that Ir--O bonds on the surface remain intact, which is consistent with the exceptional stability of IrO$_{2}$\cite{binninger2022}. A follow-up study further concluded that, for a material combining sufficient activity and stability, an OER pathway without breaking of any metal-oxygen bond should be preferred \cite{binninger2024}. Building on this realization, the first part of the present study investigates the stability of Ir--O bonds and compares them with Ru--O bonds in RuO$_{2}$ as a reference OER catalyst.\\

In another recent theoretical study, descriptors based on different thermodynamic and atomic orbital properties were evaluated to assess whether they capture the stability of OER catalysis, using IrO$_{2}$ and doped Ir-catalysts as representative examples, and applying machine learning techniques \cite{maraschin2026}. This work proposed a model for estimating the corrosion energy based on three atomic orbital descriptors, one of which reflects the Ir--O bond character through the p-band center of the oxygen ($\langle\epsilon_{2p}^\mathrm{Osite}\rangle$) \cite{maraschin2026}. The use of atomic orbital descriptors in catalysis has a long history. In their seminal work, N{\o}rskov et al. identified the d-band center as a suitable descriptor of electrocatalytic activity \cite{norskov2009}. In the present work, we follow an alternative approach, assessing the Ir--O bond character through the negative integrated Crystal Orbital Hamilton Population ($-$iCOHP), which enables the assessment of bond strength within a molecular orbital framework.\\

Previous investigations of the electronic structure of rutile IrO$_{2}$ have highlighted the critical role of metal d-orbital electrons \cite{ping2015, kahk2014}. It has been demonstrated that crystal field splitting and orbital overlap between neighbouring atoms dominate over valence electron spin-orbit coupling and electron-electron repulsion, a hierarchy that gives rise to the metallic behavior of IrO$_{2}$ \cite{ping2015,kahk2014}. Furthermore, although the M--O bond is predominantly covalent in character, an ionic contribution must not be neglected. The formal oxidation states of Ir($4+$) and O($2-$) reflect this ionic character, and an ionic description of the Ir--O bond assumes a considerable degree of electron localization on the metal and oxygen atoms, respectively \cite{ping2015}. This suggests that a complete description, of the M--O bond requires accounting for both covalent and ionic contributions.\\

Prior investigation into the electronic structure of IrO$_{2}$, using the projected density of states (pDOS), identified a notable occupation of t$_{2g}$ atomic orbitals on the Ir atom in the vicinity of the Fermi level (E$_\mathrm{F}$) \cite{ping2015}. These t$_{2g}$ orbitals contribute only weakly to the bonding states while predominantly contributing to antibonding states near E$_\mathrm{F}$, and were found to largely determine the overall electronic structure of IrO$_{2}$ \cite{ping2015}. The present work extends this perspective by adopting a molecular orbital framework that employs density functional theory (DFT). Additionally, chemical bonding analysis with the local orbital basis set towards electronic reconstruction (LOBSTER) framework is employed \cite{maintz2016}. We focus on quantifying the contribution of antibonding molecular orbitals near E$_{F}$ to the M--O bond strength. Results are compared against a detailed analysis of the electronic structure of RuO$_{2}$. In addition, through examining a range of 4d and 5d rutile oxides this study investigates the extent to which t$_{2g}$ filling of the metal atom influences the M--O bond strength by varying the number of valence electrons in the metal ($N_{\mathrm{VE,M}}$) and examining the impact on the occupation of antibonding orbitals.\\

\section{Methods}

To investigate the M--O bonding in rutile-structure MO$_{2}$ oxides, Kohn-Sham density functional theory (DFT) calculations were carried out \cite{kohn1965,kohn1996}. All simulations were performed using the Vienna Ab Initio Simulation Package (VASP) version 6.4.3, employing the projector augmented-wave (PAW) method in combination with the Perdew-Burke Ernzerhof (PBE) generalized-gradient approximation (GGA) for the exchange-correlation functional 
\cite{kresse1996,kresse1996b,blochl1994,kresse1999,perdew1996}. The study encompasses the following 4d and 5d rutile-type dioxides MO$_{2}$ (M = Mo, Ru, Rh, Pd, Ag, Ta, W, Re, Os, Ir, Pt, Au). Among these, PdO$_{2}$, AgO$_{2}$, TaO$_{2}$, WO$_{2}$, ReO$_{2}$, PtO$_{2}$ and AuO$_{2}$ represent hypothetical computational models, as either no experimental evidence for their synthesis has been reported or they are not known to adopt the exact same rutile structure as in this study was used for IrO$_{2}$ and RuO$_{2}$. Initial crystal structures for the rutile-type MO$_{2}$ materials, were obtained from the Crystallography Open Database if applicable \cite{bolzan1997,seisenbaeva2004,muller1968,baur1971, goldschmidt1926}. For the investigation of the bulk MO$_{2}$ materials, the tetragonal unit cell of the rutile structure containing two metal atoms and four oxygen atoms. Identical computational parameters were applied across all investigated binary oxides. To systematically investigate the influence of varying $N_{\mathrm{VE,M}}$, projected-augmented wave (PAW) pseudopotentials were adopted for the metal atoms to ensure a consistent treatment across all $N_{\mathrm{VE,M}}$ configurations, as provided by the VASP pseudopotential library. \cite{kresse1999}. The more noble the metal is in the 4d/5d series the higher the number of electrons in the d-orbital documented in the respective pseudopotential. For self-consistent field (SCF) calculations, a plane-wave basis cutoff of 520\,eV was employed, and Fermi-Dirac smearing with $k_\mathrm{B}T =$ 0.05\,eV was used. The D3(BJ) correction was included to account for long-range dispersion interactions \cite{grimme2010,grimme2011}. A 7 $\times$ 7 $\times$ 11 k-point grid was employed for the geometric optimizations, while a denser 17 $\times$ 17 $\times$ 21 grid was used for the subsequent single-point calculations. Geometry optimizations (including cell parameters and atom positions) were performed with a convergence threshold of -1e-2\,eV / \AA{}.

\subsection{Electronic Structure Analysis}
 As a starting point for the analysis of the electronic structure, the density of states (DOS) near $E_\mathrm{F}$ was calculated for each material. The DOS is obtained as a summation over all Kohn-Sham (KS) eigenvalues ($\epsilon_{i}$) which are calculated through DFT simulations, 

\begin{equation}
    \mathrm{DOS} \left(E\right) = \mathlarger{\sum}_{i} \delta \left( E - \epsilon_{i} \right).
\end{equation}

\noindent The DOS can be projected onto specific atomic orbitals $\mu$ resulting in the pDOS. The pDOS reveals the contribution of each atomic orbital to the KS eigenvalues, 

\begin{equation}
    \mathrm{pDOS}_{\mu}\left(E\right) =  \mathlarger{\sum}_{i} \left| c_{i\mu}\right|^{2} \delta \left( E - \epsilon_{i} \right).
    \label{pdos}
\end{equation}

\noindent The additional information is obtained by including the orbital coefficients $c_{i\mu}$ as weighting factors. $c_{i\mu}$ is defined through the projection of the plane-wave Kohn-Sham eigenstates $\ket{\psi_{i}}$ onto the atomic orbitals $\ket{\phi_{i}}$,

\begin{equation}
    \ket{\psi_{i}} = \sum_{\mu}c_{i\mu}\ket{\phi_{\mu}}
    \label{wavefunction}
\end{equation}

\noindent where $c_{i\mu}$ denotes the $\mu$-th component of the eigenvector for state $i$ in the respective basis. The pDOS provides a decomposition of the electronic structure in terms of atomic orbitals. However, to investigate the character of the M--O bonds in greater detail, the electronic structure must be analyzed in terms of \emph{molecular orbitals}, or, in the case of periodic structures, \emph{crystal orbitals}. 

\subsection{Covalent Bond Strength}

The strength of a covalent bond can be assessed using the crystal orbital Hamilton population (COHP) analysis. The COHP analysis was performed using the  LOBSTER software \cite{maintz2016}. LOBSTER reconstructs chemical bonding information from plane-wave DFT calculations by projecting the plane-wave-based wavefunctions onto a localized atomic orbital basis. Based on this projection, LOBSTER reconstructs the Kohn-Sham Hamiltonian matrix elements in the local atomic orbital basis. This reconstruction corresponds to a re-expression of the delocalized plane-wave Kohn-Sham eigenstates, obtained from the DFT calculations, in terms of a localized atomic orbital basis set. The resulting reconstructed matrices enable the identification of bonding ($\mathrm{COHP} < 0$) and antibonding ($\mathrm{COHP} > 0$) contributions to a chemical bond, through the computation of bond descriptors such as the COHP \cite{dronskowski1993}. A more positive COHP values indicates stronger antibonding interactions, which correspond to a less covalent M--O bond.\\

To understand the COHP concept on a fundamental level, it is instructive to consider the bandstructure energy $E_\mathrm{BS}$, defined as,

\begin{equation}
    E_\mathrm{BS} = \mathlarger{\sum}_{i} f_{i} \epsilon_{i}
    \label{band_structure_energy}
\end{equation}

\noindent where $f_{i}$ denotes the occupation of state $i$ and $\epsilon_{i}$ the corresponding eigenvalue. Using the expansion of the Kohn-Sham eigenstates in the local atomic orbital basis (Equation \ref{wavefunction}) and applying the Kohn-Sham eigenvalue equation, $\epsilon_{i}$ can be expressed as,

\begin{equation}
    \epsilon_{i} = \mathlarger{\sum}_{\mu}\mathlarger{\sum}_{\nu}c_{i\mu}^{*}c_{i\nu}H^\mathrm{KS}_{\mu\nu}.
\end{equation}

\noindent Inserting this into the expression for $E_\mathrm{BS}$ yields, 

\begin{equation}
    E_\mathrm{BS} = \mathlarger{\sum}_{i} f_{i} \mathlarger{\sum}_{\mu}\mathlarger{\sum}_{\nu}c_{i\mu}^{*}c_{i\nu}H^\mathrm{KS}_{\mu\nu}.
\end{equation}

To obtain an energy-resolved representation, a Dirac delta function is introduced to select contributions at each energy level $\epsilon_{i}$,

\begin{equation}
    E_\mathrm{BS} = \int_{-\infty}^{E_\mathrm{F}}dE \mathlarger{\sum}_{i}\mathlarger{\sum}_{\mu}\mathlarger{\sum}_{\nu}\mathrm{Re}\left[c_{i\mu}^{*}c_{i\nu}H^\mathrm{KS}_{\mu\nu}\,\delta \left( E - \epsilon_{i} \right)\right]
\end{equation}

\noindent Since the physically meaningful energy contribution must be real-valued, the real part of the integrand is retained. The resulting integrand defines the COHP density between atom $A$ and $B$\cite{dronskowski1993,deringer2011}, 

\begin{equation}
    \mathrm{COHP}_{AB}\left(E\right) = \mathlarger{\sum}_{i}\mathlarger{\sum}_{\mu \in A}\mathlarger{\sum}_{\nu \in B} H_{\mu\nu}^{\mathrm{KS}} \mathrm{Re}\left[c_{i\mu}^{*}c_{i\nu} \, \delta \left( E - \epsilon_{i} \right)\right].
    \label{cohp}
\end{equation}
\noindent Here, to investigate the M--O bonding characteristics in the MO$_{2}$ systems, chemical bonding analysis was performed using LOBSTER with the pbeVaspFit2015 basis set \cite{nelson2020}. The basis functions were selected to incorporate the atomic orbitals critical to M--O interactions. Specifically, the oxygen 2s and 2p, and metal 5d and 6s, or 4d and 5s states were included for the 5d and 4d transition metal series, respectively. The COHP was calculated across the entire energy range of the KS DOS to evaluate all bonding and antibonding contributions. To quantify the overall covalent bond strength and assess the stability of the M--O interactions, the COHP was integrated from the lowest occupied KS energy state up to the Fermi energy, $E_\mathrm{F}$, yielding the iCOHP. The negative iCOHP, denoted as $-$iCOHP, is used throughout this work, as it provides a more intuitive representation of the covalent bond strength: a more positive $-$iCOHP value corresponds to a stronger covalent bond. This analysis specifically considered M--O interactions within the primary coordination sphere of the  rutile MO$_{2}$ materials \cite{bolzan1997,park2025}.\\
 
The $-$iCOHP is compared with three other characteristics: the M--O bond length, d-band center, and valence electron ($N_{\mathrm{VE,M}}$) count of the respective metal. Additionally, the orbital overlap between the metal d-orbital and the oxygen p-orbital is also included as it is one of the key factors determining the covalent bond strength \cite{ping2015}. The results showing the relationship between $-$iCOHP and the d-band center or the orbital overlap are provided in the Supplementary Information (SI) in Figures S11 and S12, as both of these descriptors are grounded in the atomic orbital picture rather than molecular orbital framework adopted in the main text. It is generally assumed that shorter bond lengths correspond to stronger orbital overlap between metal and oxygen atoms, increasing electron sharing and covalency. The energetic center of the metal d- or oxygen p-orbital up to $E_\mathrm{F}$ is calculated as,

 \begin{equation}
    \varepsilon_\mathrm{d/p}^\mathrm{occ.} = \frac{\int_{-\infty}^{E_\mathrm{F}}E \cdot \mathrm{pDOS}_\mathrm{d/p}(E)\mathrm{d}E}{\int_{-\infty}^{E_\mathrm{F}}\mathrm{pDOS}_\mathrm{d/p}(E)\mathrm{d}E}.
 \end{equation} 
 
 \noindent For the $N_{\mathrm{VE,M}}$ count, we  generally expect that an increased d-orbital occupancy correlates with weaker M--O covalency due to enlarged antibonding orbital population. It is tested whether orbital overlap can be used as a covalent bond strength descriptor. Therefore, the energetic difference between the occupied d-band center and the p-band center ($\varepsilon_\mathrm{d,M}^\mathrm{occ.} - \varepsilon_\mathrm{p,O}^\mathrm{occ.}$) was calculated.\\ 

$\varepsilon_\mathrm{d,M}^\mathrm{occ.} - \varepsilon_\mathrm{p,O}^\mathrm{occ.}$ was determined using the VASPKIT software package \cite{wang2021}. For the present analysis, only occupied states were considered. Therefore, the metal d- or oxygen p-orbital was integrated from its lowest energy up to $E_\mathrm{F}$. Positive values of $\varepsilon_\mathrm{d,M}^\mathrm{occ.} - \varepsilon_\mathrm{p,O}^\mathrm{occ.}$ indicate that metal d-orbitals lie at higher energy relative to oxygen p-orbitals, corresponding to reduced hybridization between metal d- and oxygen p-states. Values of $\varepsilon_\mathrm{d,M}^\mathrm{occ.} - \varepsilon_\mathrm{p,O}^\mathrm{occ.}$ close to zero reflect strong orbital overlap and enhanced hybridization, which indicates a more covalent M--O character with greater electron sharing between metal and oxygen atoms. When metal d-orbitals lie substantially lower in energy than oxygen p-orbitals, oxygen p-states are less likely to form bonds with metal d-orbitals due to its high energetic difference and instead preferentially interact with metal s- or p-orbitals since these are energetically more favorable \cite{mulliken1950, krapp2006}.

 \subsection{Ionic lattice energy}
The ionic lattice energy is an expression for the ionic bond strength. It can be defined as the energy required to separate a pair of oppositely charged ions. While in covalent bonds electrons are shared between the M atom and the O atom, for the ionic lattice energy the electrons remain localized on the individual atomic sites \cite{bader1965}.\\

The ionic lattice energy can be assessed through the Madelung Energy $E_\mathrm{M}$, which describes the total electrostatic energy of an infinite periodic ionic crystal. In the LOBSTER software, $E_\mathrm{M}$ is computed by using the DFT-derived charges $q_{i}$ in an Ewald summation. The Madelung energy is obtained from LOBSTER as,
\begin{equation}
    E_\mathrm{M} = \frac{1}{2} \sum_{i}q_{i}\phi_{i} \ .
\end{equation}

\noindent $E_\mathrm{M}$ obtained from LOBSTER is normalized per formula unit. In the present work, $E_\mathrm{M}$ was subsequently normalized per bond to obtain a quantity directly comparable to the covalent bond strength descriptor $-$iCOHP.\\

As a descriptor for \emph{ionicity}, i.e., the degree of ionic character, the ratio $\chi_\mathrm{bond}$, which quantifies the extent of actual charge transfer between atoms relative to the idealised ionic picture, is used,
\begin{equation}
    \chi_\mathrm{bond} = \frac{q_\mathrm{i}}{q_\mathrm{formal}}.
    \label{ionicity}
\end{equation}

 \noindent $q_\mathrm{i}$ represents the estimated ionic charge in the material. In this study, it is computed as the Löwdin charge using LOBSTER \cite{maintz2013,maintz2016}. $q_\mathrm{formal}$ is the integer charge expected based on formal oxidation states. For all MO$_{2}$ materials examined here, formal oxidation states of $-2$ for oxygen and $+4$ for metal atoms are assumed. When $\chi_\mathrm{bond}$ approaches a value of unity, the effective charge is close to the formal charge, indicating nearly ideal ionic behaviour and, thus, a maximal ionicity. Values closer to zero signify that the effective ionic charge is much smaller than the formal valency, pointing to substantial covalent character in which electrons are shared rather than transferred between the constituent atoms \cite{dronskowski2023}.\\

\section{Results}

\begin{table}[!t]
\caption{DFT-optimized lattice parameters for rutile MO$_{2}$ structures of the 4d series. Values in parentheses indicate the experimental literature values \cite{bolzan1997,seisenbaeva2004,muller1968,baur1971}. Materials marked with an asterisk (*) are to be considered as artificial phases, since no experimental rutile-structure counterparts are known.}
\label{geo_opt_4d}
\resizebox{\textwidth}{!}{%
\begin{tabular}{cccccc}
\hline
\begin{tabular}[c]{*{6}{>{\centering\arraybackslash}p{0.167\textwidth}}}Lattice \\ Parameters\\ (\AA{} / $\circ$)\end{tabular} & MoO$_{2}$       & RuO$_{2}$        & RhO$_{2}$       & PdO$_{2}$ & AgO$_{2}$* \\ \hline
$a$                                                                                  & 4.95 (4.85 \cite{seisenbaeva2004})   & 4.50 (4.50 \cite{bolzan1997}) & 4.50 (4.49 \cite{muller1968})   & 4.53 (4.52\cite{shaplygin1978})     & 4.60      \\
$b$                                                                                  & 4.95 (4.85 \cite{seisenbaeva2004})   & 4.50 (4.50 \cite{bolzan1997}) & 4.50 (4.49 \cite{muller1968})   & 4.53 (4.52\cite{shaplygin1978})     & 4.60      \\
$c$                                                                                  & 2.72 (2.81 \cite{seisenbaeva2004})   & 3.11 (3.10 \cite{bolzan1997})    & 3.10 (3.09 \cite{muller1968})  & 3.18 (3.15\cite{shaplygin1978})     & 3.34      \\ 
$\alpha = \beta = \gamma$                                                                             & 90  (90 \cite{seisenbaeva2004}) & 90  (90 \cite{bolzan1997})  & 90  (90 \cite{muller1968}) & 90 (90 \cite{shaplygin1978})        & 90        \\ \hline
\end{tabular}%
}
\end{table}

\begin{table}[!t]
\caption{DFT-optimized lattice parameters for rutile MO$_{2}$ structures of the 5d series. Values in parentheses indicate the experimental literature values \cite{bolzan1997, goldschmidt1926}. Materials marked with an asterisk (*) are to be considered as artificial phases, since no experimental rutile-structure counterparts are known. Materials marked with two asterisks (**) indicate that these MO$_{2}$ oxides exist but either not in a rutile structure or not the same space group as considered for IrO$_{2}$ and RuO$_{2}$ in this study \cite{khitrova1968, bolzan1995,wyckoff1963}. }
\label{geo_opt_5d}
\resizebox{\textwidth}{!}{%
\begin{tabular}{cccccccc}
\hline
\makecell{Lattice \\ Parameters \\ (\AA{} / $\circ$)} & TaO$_{2}$ & WO$_{2}$** & ReO$_{2}$** & OsO$_{2}$        & IrO$_{2}$        & PtO$_{2}$       & AuO$_{2}$* \\ \hline
$a$  & 4.98 (4.71 \cite{schonberg1954}) & 4.62 & 4.93 & 4.50 (4.52\cite{goldschmidt1926}) & 4.52 (4.52 \cite{bolzan1997}) & 4.56 (4.49 \cite{fernandez1984}) & 4.60 \\
$b$  & 4.98 (4.71 \cite{schonberg1954}) & 4.62 & 4.93 & 4.50 (4.52 \cite{goldschmidt1926}) & 4.52 (4.52 \cite{bolzan1997}) & 4.56 (4.49 \cite{fernandez1984}) & 4.60 \\
$c$  & 2.86 (3.07 \cite{schonberg1954}) & 3.25 & 2.70 & 3.21 (3.20 \cite{goldschmidt1926}) & 3.18 (3.16 \cite{bolzan1997})    & 3.22 (3.13 \cite{fernandez1984}) & 3.40 \\
$\alpha = \beta = \gamma$ & 90 (90 \cite{schonberg1954}) & 90 & 90 & 90  (90 \cite{goldschmidt1926})  & 90  (90 \cite{bolzan1997})  & 90 (90 \cite{fernandez1984}) & 90 \\ \hline
\end{tabular}%
}
\end{table}

Geometry optimizations were performed for all rutile MO$_{2}$ materials. Table \ref{geo_opt_4d} (MoO$_{2}$, RuO$_{2}$, RhO$_{2}$, PdO$_{2}$, and AgO$_{2}$)and Table \ref{geo_opt_5d} (TaO$_{2}$, WO$_{2}$, ReO$_{2}$, OsO$_{2}$, IrO$_{2}$, PtO$_{2}$, and AuO$_{2}$) list the optimized lattice parameters (unit cell lengths $a$, $b$, $c$, and angle $\alpha = \beta = \gamma$) along with experimental literature values given in parentheses \cite{bolzan1997,seisenbaeva2004,muller1968,baur1971,goldschmidt1926}. Materials marked with an asterisk (*) indicate that no comparable rutile structure for these specific compositions was found in the literature, and the inital structure for geometric optimization was therefore taken from the optimized IrO$_{2}$ cell. MO$_{2}$ materials marked with (**) have been reported in experimental studies, however, since these oxides adopt a crystal structure with a space group different from the rutile P42/mnm space group considered in this work, their experimental lattice parameters were not included in the comparison \cite{khitrova1968,bolzan1995,wyckoff1963}.\\

The geometrically optimized lattice parameters are in good agreement with the experimental values reported in the literature for all studied metal oxide materials. The deviations from experiment are negligibly small across the entire set, confirming that well-converged structural models were obtained for all systems. One exception is TaO$_{2}$ where geometrical optimization results in minimal higher deviation from literature.

\subsection{Electronic Structure Analysis}

\begin{figure}[!t]
  \centering
  \includegraphics[width=1\linewidth]{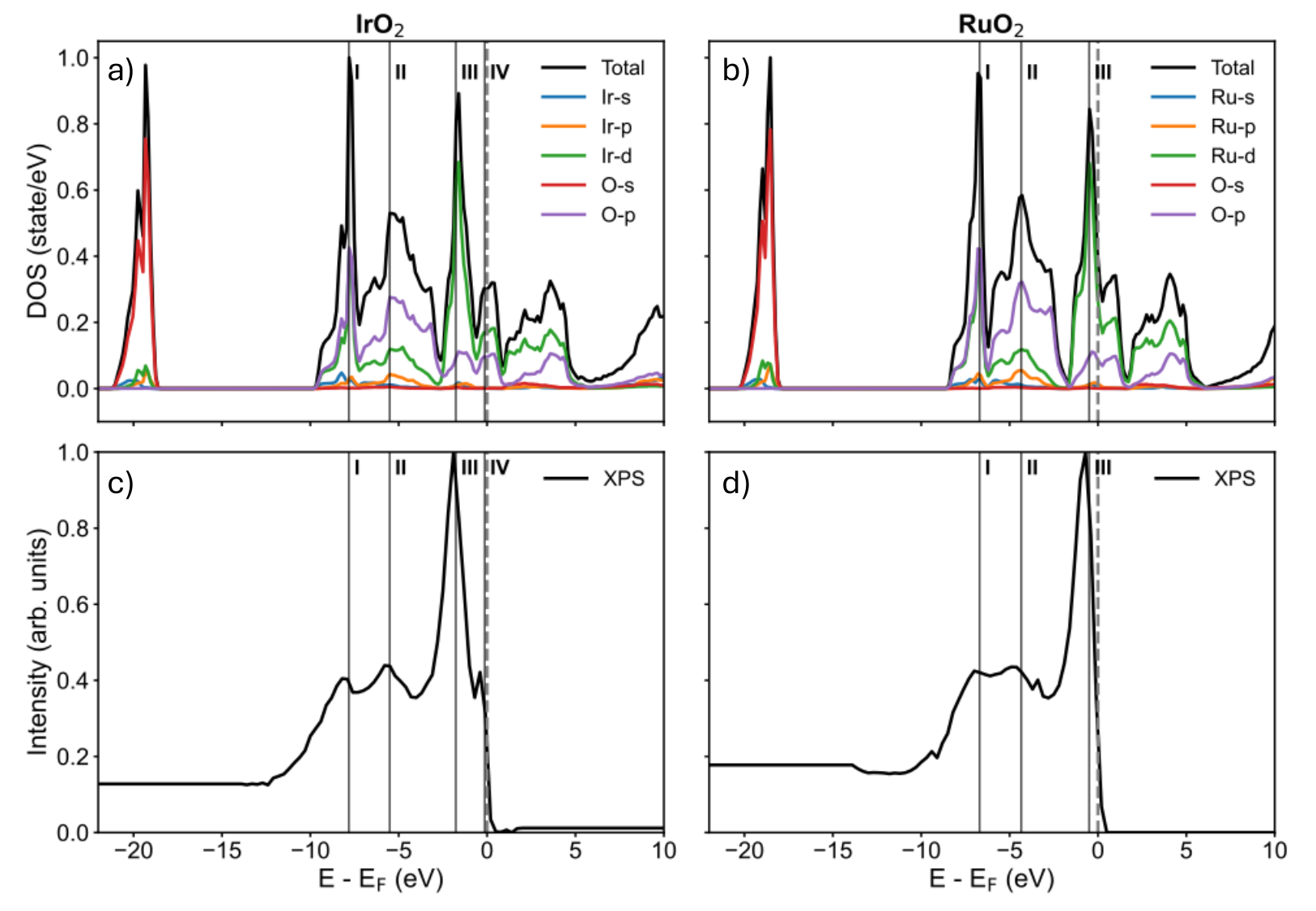}
\caption{Top: Calculated total DOS and pDOS for rutile IrO$_{2}$ (a) and RuO$_{2}$ (b) in the vicinity of $E_\mathrm{F}$ (vertical dashed-line), presented over an identical energy range for direct comparison. The pDOS resolves contributions from metal s, p, d orbitals and oxygen s, p orbital. Bottom: XPS valence-band spectra of bulk IrO$_{2}$ (c) and RuO$_{2}$ (d) as reference for comparison to the calculated DOS, aligned on the same energy axes. XPS reference data taken from Kahk et al. \cite{kahk2014} and Riga et al. \cite{riga1977} for IrO$_{2}$ and RuO$_{2}$, respectively. Vertical lines mark the peak positions in the DOS and XPS spectra and are numbered accordingly.}
\label{pdos_xps}
\end{figure}

Before analysing the trends in M--O bond strength across the 4d and 5d series of rutile-type oxides, the electronic structures of IrO$_{2}$ and RuO$_{2}$ are discussed in more detail, as these are the primary materials of interest in the context of OER electrocatalysis. The pDOS for the remaining 4d and 5d series of rutile-type oxides can be found in the SI from Figure S1 to S10. Figure \ref{pdos_xps} (a) and (b) presents the total DOS and pDOS, resolving metal s, p, d orbitals and oxygen s, p orbitals for each material. The DOS profiles of IrO$_{2}$ and RuO$_{2}$ are very similar, but shifted corresponding to the respective Fermi energies. This relative shift reflects the higher d-electron count ($N_{\mathrm{VE,M}}$) and band filling in IrO$_{2}$ compared to RuO$_{2}$, with 9 v$N_{\mathrm{VE,M}}$ of iridium versus 8 $N_{\mathrm{VE,M}}$ for ruthenium. Published data from valence-band X-ray photoelectron spectroscopy (XPS) measurements by Kahk et al. \cite{kahk2014} and Riga et al. \cite{riga1977} are shown in Figure \ref{pdos_xps} (c) and (d) for comparison, and demonstrate very good agreement with DOS calculated at the GGA level.\\

\begin{figure}[!t]
  \centering
  \includegraphics[width=1\linewidth]{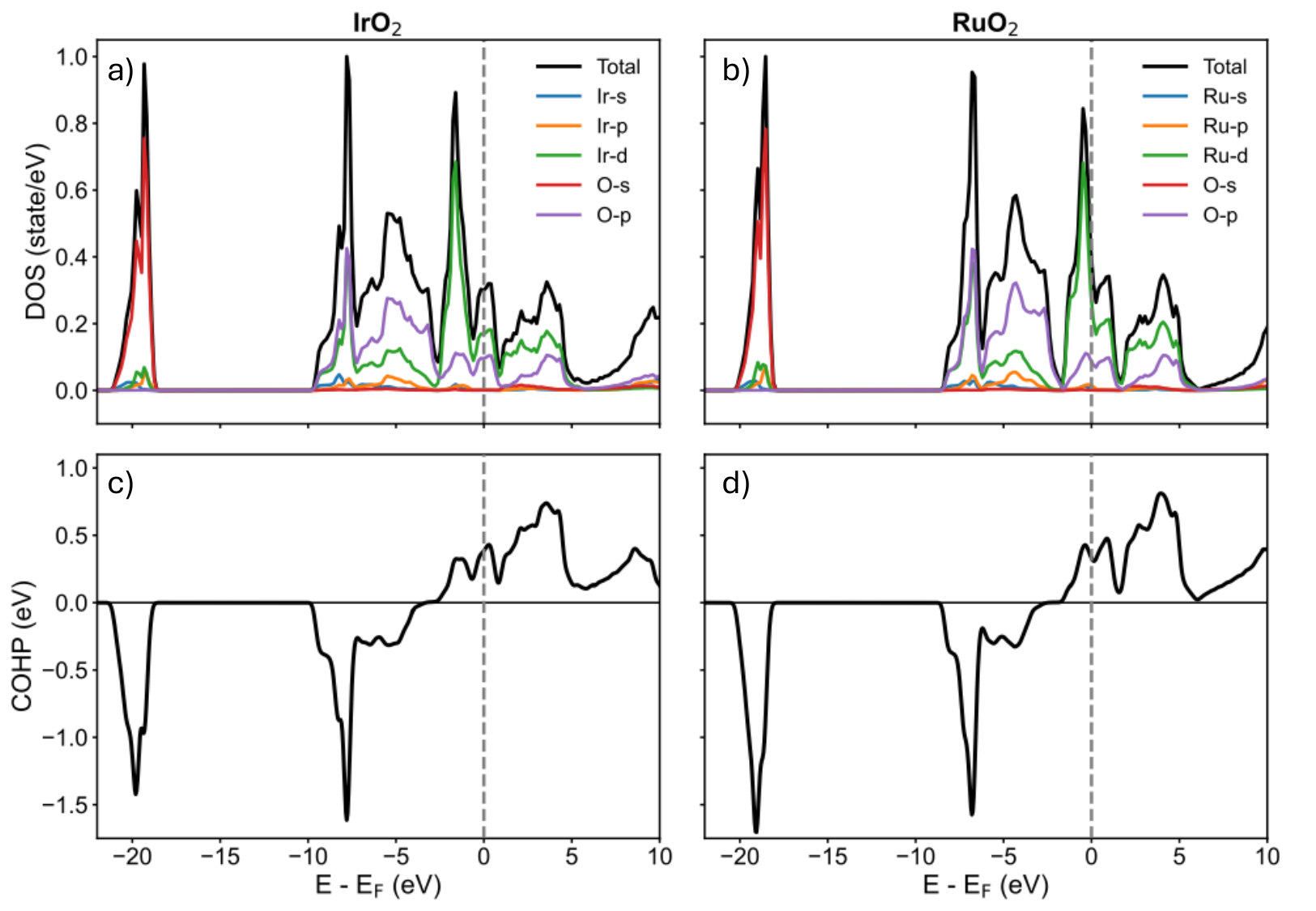}
\caption{Bottom: COHP curves for rutile IrO$_{2}$ (c) and RuO$_{2}$ (d), showing the total M--O contribution. The total DOS and pDOS, already shown in Figure~\ref{pdos_xps}, are shown again for comparison (a) and b)). All energy axes are referenced to $E_\mathrm{F}$, and identical axis limits and tick marks are used across panels to facilitate comparison between features in DOS/pDOS and COHP.}
\label{pdos_cohp}
\end{figure}

To further compare the M--O bonding in IrO$_{2}$ and RuO$_{2}$, Figure \ref{pdos_cohp} (c) and (d) shows the respectiveCOHP for both materials. In both systems, strongly bonding molecular orbitals near $-20$\,eV (versus $E_\mathrm{F}$) are observed. Based on the pDOS reproduced in Figure \ref{pdos_cohp} (a) and (b), these bonding orbitals are dominated by O s-orbitals, with minor contributions from metal p- and d-orbitals. Additional bonding contributions occur at approximately $-8$\,eV for IrO$_{2}$ and $-7$\,eV for RuO$_{2}$, primarly arising from O-p--M-d interactions. These bonding contributions stabilize the M--O bonds in IrO$_{2}$ and RuO$_{2}$. In IrO$_{2}$ around $-3.3$\,eV and RuO$_{2}$ around $-2.5$\,eV, the COHP switches to positive values, indicating antibonding character and corresponding bond destabilization. This onset coincides with increased M-d contributions relative to O-p in the pDOS. It is noted that the energetic regions in the DOS identified as bonding or antibonding on the basis of the COHP are in agreement with previous findings, e.g., by Ping et al. \cite{ping2015}. In that study, the metal d- and oxygen p-orbitals near E$_\mathrm{F}$ were classified according to their spatial orientation relative to one another, exhibiting either bonding character in the form of $\sigma$ or $\pi$ molecular orbitals, or antibonding character in the form of $\sigma^{*}$ or $\pi^{*}$ molecular orbitals.  Integrating the total COHP yields almost identical per-bond values of $-3.48$\,eV for IrO$_{2}$ and $-3.54$\,eV for RuO$_{2}$. Interestingly, based on these iCOHP values, the covalent M--O bonds appear to be of comparable strength in bulk RuO$_{2}$ relative to IrO$_{2}$. This finding stands in contrast to the conclusions drawn by Tang et al., who reported that IrO$_{2}$ exhibits more covalent bonding characteristics associated with greater electron filling in the Ir--O bonds \cite{tang2025}. The discrepancy may be attributed to the choice of pseudopotentials: in the present study, pseudopotentials for all 4d and 5d materials were selected to ensure mutual consistency, such that variations in $N_{\mathrm{VE,M}}$, across the molecular orbitals can be systematically studied.\\

It is observed that the COHP curves for IrO$_{2}$ and RuO$_{2}$ exhibit a nearly identical structure, differing primarily by a slight shift versus the respective Fermi energy, consistent with the observations for the corresponding DOS curves discussed before, which are attributed to the different $N_{\mathrm{VE,M}}$ numbers of iridium (9 $N_{\mathrm{VE,M}}$) and ruthenium (8 $N_{\mathrm{VE,M}}$). The extra electron of iridium results in a slightly higher filling of antibonding orbitals in IrO$_{2}$, which is expected to weaken the respective M--O bonds. From this perspective, the essentially equal bond strength in IrO$_{2}$ and RuO$_{2}$ as established by the iCOHP is surprising and indicates another competing factor at play in determining the relative stability of the M--O bonds in these materials. \\

\subsection{Covalent Bond Strength}

\begin{table}[!t]
\caption{Comparison of different descriptors for the covalent bond strength (number of antibonding electrons, $-$iCOHP, $\varepsilon_\mathrm{d,M}^\mathrm{occ.} - \varepsilon_\mathrm{p,O}^\mathrm{occ.}$, and $l$ across various rutile-type MO$_{2}$ materials with different numbers of valence electrons of the transition metals. Materials marked with asterisk (*) indicate that these materials are not known experimentally. Materials marked with two asterisks (**) indicate that these metal oxides exist but not in the rutile structure \cite{khitrova1968, bolzan1995,wyckoff1963}.}
\label{covalent_bond_strength_table}
\resizebox{\textwidth}{!}{%
\begin{tabular}{cccccc}
\hline
Material    & $N_{\mathrm{VE,M}}$ & $N_{\mathrm{ab}}$ & \begin{tabular}[c]{@{}c@{}}$-$iCOHP  \\ (eV)\end{tabular} & \begin{tabular}[c]{@{}c@{}}$\varepsilon_\mathrm{d,M}^\mathrm{occ.} - \varepsilon_\mathrm{p,O}^\mathrm{occ.}$\\ (eV)\end{tabular} & \begin{tabular}[c]{@{}c@{}}$l$\\ (\AA{})\end{tabular} \\ \hline
MoO$_{2}$   & 6                   & 1.99              & 4.04                                                      & -0.44                                                                                                                            & 2.02                                                                   \\
RuO$_{2}$   & 8                   & 4.02              & 3.54                                                      & 0.49                                                                                                                             & 1.97                                                                   \\
RhO$_{2}$   & 9                   & 6.03              & 3.17                                                      & 0.08                                                                                                                             & 1.97                                                                   \\
PdO$_{2}$  & 10                  & 7.62              & 2.65                                                      & -0.73                                                                                                                            & 2.00                                                                   \\
AgO$_{2}$*  & 11                  & 8.56              & 1.93                                                      & -1.69                                                                                                                            & 2.06                                                                   \\
TaO$_{2}$ & 5                   & 0.86              & 4.26                                                      & -1.65                                                                                                                            & 2.06                                                                   \\
WO$_{2}$**  & 6                   & 1.85              & 4.06                                                      & -0.62                                                                                                                            & 2.04                                                                   \\
ReO$_{2}$** & 7                   & 2.93              & 3.99                                                      & 0.30                                                                                                                             & 2.01                                                                   \\
OsO$_{2}$   & 8                   & 3.65              & 3.83                                                      & 0.57                                                                                                                             & 2.00                                                                   \\
IrO$_{2}$   & 9                   & 5.41              & 3.48                                                      & 0.71                                                                                                                             & 1.99                                                                   \\
PtO$_{2}$ & 10                  & 7.32              & 2.96                                                      & -0.13                                                                                                                            & 1.95                                                                   \\
AuO$_{2}$*  & 11                  & 8.74              & 2.23                                                      & -1.30                                                                                                                            & 2.07                                                                   \\ \hline
\end{tabular}%
}
\end{table}

In search of this other factor, the $-$iCOHP was evaluated for additional rutile-type MO$_{2}$ oxides of transition metals adjacent to Ir and Ru in the periodic table (with M=Mo, Rh, Pd, Ag, Ta, W, Re, Os, Pt, Au, see Tables \ref{geo_opt_4d} and \ref{geo_opt_5d}). The respective pDOS and COHP for the added 4d and 5d metal oxides are shown in the SI (Figure S1 - S10). Table \ref{covalent_bond_strength_table} displays the various bonding descriptors calculated for all oxides. The number of valence electrons of the transition metal ($N_{\mathrm{VE},M}$), the number of antibonding electrons ($N_{\mathrm{ab}}$), the $-$iCOHP, the bond length $l$, and the difference between the d-band center of the metal and the p-band center of the oxygen atoms ($\varepsilon_\mathrm{d,M}^\mathrm{occ.} - \varepsilon_\mathrm{p,O}^\mathrm{occ.}$). The latter property was included to examine the relationship between orbital overlap and covalent bond strength. The relation between $-$iCOHP and orbital overlap is shown in Figure S12. Obviously, the number of antibonding electrons increases with $N_{\mathrm{VE,M}}$, as already established by the comparison between IrO$_{2}$ and RuO$_{2}$ in Figure \ref{pdos_cohp}. This shows that across all materials considered, the Fermi level is located in the antibonding band and the increasing number of metal valence electrons thus leads to a systematic increase in the filling of antibonding molecular orbitals. 

\begin{figure}[!t]
  \centering
  \includegraphics[width=1\linewidth]{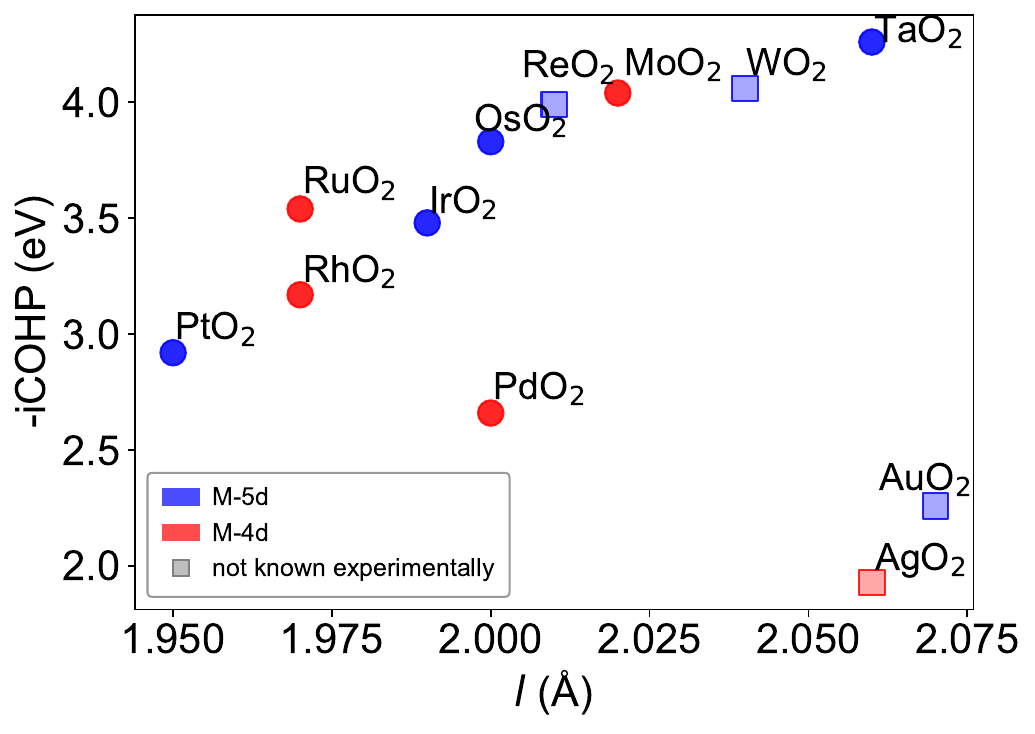}
\caption{$-$iCOHP versus $l$ of the DFT-relaxed structures for the rutile MO$_{2}$ materials. Data for 5d-metals are shown in blue, and data for 4d-metals are shown in red. The dashed circles around a data point indicate that no experimental structural data was found for the respective material, which therefore must be treated as artificial model materials.}
\label{icohp_bond_length}
\end{figure}

To examine whether purely structural properties can serve as viable descriptors, the M--O bond length $l$ was investigated. Since all studied rutile-type metal oxides adopt an octahedral coordination environment, $l$ was computed as the weighted average of the four equatorial and two axial M--O bond lengths. Figure \ref{icohp_bond_length} shows $-$iCOHP plotted against $l$. Across the rutile MO$_2$ series considered here, the average bond lengths of the DFT-relaxed structures were rather similar, spanning a range of only about 0.1\,\AA{} between the shortest and longest values within this series. 
 Overall, Figure \ref{icohp_bond_length} reveals no clear correlation between $-$iCOHP and $l$, which is likely due to the minor spread in the M--O bond lengths among the rutile MO$_{2}$ materials considered here.\\

\begin{figure}[!t]
  \centering
  \includegraphics[width=1\linewidth]{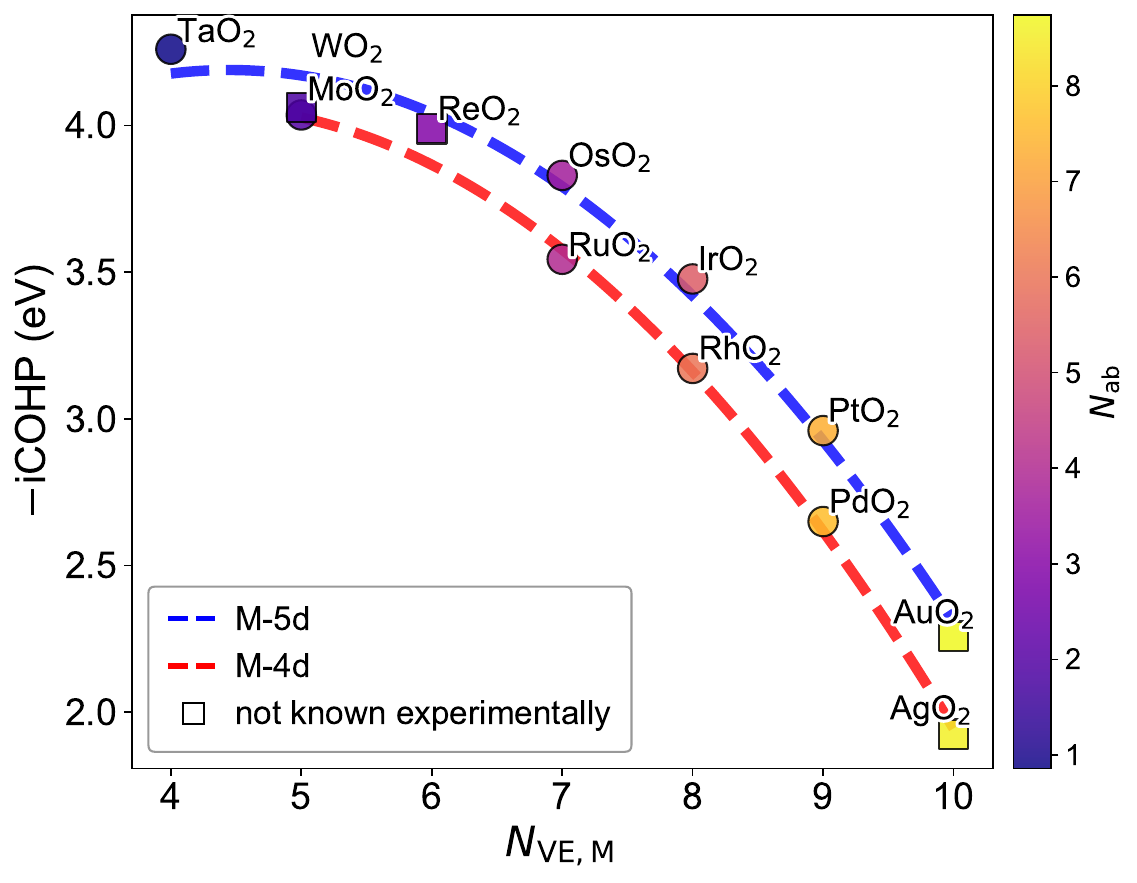}
\caption{$-$iCOHP as a function of the number of $N_{\mathrm{VE,M}}$ in the metal d-orbital. The trends were fitted separately for the 4d (red) and 5d (blue) series using a quadratic fitting function. Data points are color-coded by the number of states in the antibonding $\pi^{*}$ molecular orbital. The dashed circles indicate the artificial materials where no experimental counterparts are known.}
\label{icohp_valence_electrons}
\end{figure}

In Figure \ref{icohp_valence_electrons},  $-$iCOHP of the M--O bonds in MO$_{2}$ oxides is plotted as a function of the number of $N_{\mathrm{VE,M}}$. The data points of the materials are color-coded according to $N_{\mathrm{ab}}$ in Table \ref{covalent_bond_strength_table}, which represent the number of electrons in the $\pi^{*}$ molecular orbital. The data points reveal two clear trends within the 4d and 5d transition metal series, which were well described by quadratic fits. The 5d fit yielded,

\begin{equation}
    -\mathrm{iCOHP}(\mathrm{M^{5d}O_2}) = -0.061\,\mathrm{eV} \,\left(N_{\mathrm{VE,M}}-4.45\right)^{2} + 4.19\,\mathrm{eV} \ ,
\end{equation}
\noindent and the 4d fit

\begin{equation}
    -\mathrm{iCOHP}(\mathrm{M^{4d}O_2}) = -0.064\,\mathrm{eV} \,\left(N_{\mathrm{VE,M}}-4.24\right)^{2} + 4.06\,\mathrm{eV} \ ,
\end{equation}
The quadratic trends in 4d and 5d transition metals have a similar curvature and exhibit their maximal covalent bond strength $N_{\mathrm{VE,M}}$ of $N_\mathrm{VE,M}\approx4.3$, which is essentially identical for both series. Importantly, an overall offset in $-$iCOHP of about 0.13\,eV separates the 4d and 5d series. This likely arises from the greater spatial extent of the 5d orbitals, and the corresponding enhanced overlap with O-2p orbitals that results in stronger covalent bonds in MO$_{2}$ materials of the 5d series. This offset also explains why IrO$_{2}$ and RuO$_{2}$ exhibit similar covalent bond strength despite their different d-electron counts: going from RuO$_{2}$ to IrO$_{2}$, the change from the 4d to the 5d manifold leads to an enhancement of the M--O bond strength, which is compensated by the increased population of antibonding orbitals due the extra valence electron of Ir in comparison to Ru. Coincidentally, these two opposing effects cancel each other, resulting in almost identical $-$iCOHP values and, thus, strength of the covalent M--O bonds for both materials, as seen in Figure~\ref{icohp_valence_electrons}. This observation was made possible by the systematic selection of pseudopotentials described in the Methods section, where pseudopotentials for the 4d and 5d metals were chosen to consistently capture the influence of $N_\mathrm{VE,M}$ on $N_{\mathrm{ab}}$. For some metals we therefore employed pseudopotentials from the VASP library which deviate from their default recommendations. Based on the trends observed in Figure~\ref{icohp_valence_electrons} we consider our results to be well-founded. One reason for the discrepancy between our study and the findings of Tang et al. \cite{tang2025} may be the use of different pseudopotentials for the respective metal atom. This assumption cannot be verified as the specific choice of pseudopotentials was not documented in the study by Tang et al \cite{tang2025}.\\

The same factors explain the trends across the 4d and 5d series seen in Figure \ref{icohp_valence_electrons}, with a general decrease in covalent bond strength ($-$iCOHP) with increasing number of $N_{\mathrm{VE,M}}$ due to enhanced occupation of antibonding states. It is worth noting that closely analogous trends ave recently been reported for Y$_{2}$M$_{3}$Si$_{5}$ (M = Mn-Cu, Tc-Pd, Re-Pt) intermetallic compounds by Palla et al. \cite{palla2025}. 

\subsection{Ionic lattice energy}

\begin{table}[!t]
\caption{Comparison of the ionicity and ionic lattice energy descriptors (Löwdin charges, $\chi_\mathrm{bond}$, and $E_\mathrm{M}$) across various MO$_{2}$ materials of the 4d series. Materials marked with asterisk (*) indicate that these materials are not known experimentally.}
\label{ionic_bond_strength_4d}
\resizebox{\textwidth}{!}{%
\begin{tabular}{cccccc}
\hline
Material                                                     & MoO$_{2}$                                                 & RuO$_{2}$                                                 & RhO$_{2}$                                                 & PdO$_{2}$                                                   & AgO$_{2}$*                                                   \\ \hline
Löwdin charges ($e$)                                         & \begin{tabular}[c]{@{}c@{}}Mo: 1.22\\ O: $-0.61$\end{tabular} & \begin{tabular}[c]{@{}c@{}}Ru: 1.22\\ O: $-0.61$\end{tabular} & \begin{tabular}[c]{@{}c@{}}Rh: 1.28\\ O: $-0.64$\end{tabular} & \begin{tabular}[c]{@{}c@{}}Pd: 1.30\\ O: $-0.65$\end{tabular} & \begin{tabular}[c]{@{}c@{}}Ag: 1.50\\ O: $-0.75$\end{tabular} \\
$\chi_\mathrm{bond}$ & 0.30                                                        & 0.31                                                        & 0.32                                                        & 0.33                                                        & 0.38                                                        \\
$-E_\mathrm{M}$                                            & 2.06                                                        & 2.20                                                        & 2.41                                                        & 2.45                                                        & 3.14                                                        \\ \hline
\end{tabular}%
}
\end{table}

\begin{table}[!t]
\caption{Comparison of the ionicity and ionic lattice energy descriptors (Löwdin charges, $\chi_\mathrm{bond}$, and $E_\mathrm{M}$) across various MO$_{2}$ materials of the 5d series. Materials marked with asterisk (*) indicate that these materials are not known experimentally. Materials marked with two asterisks (**) indicate that these metal oxides exist but not in the rutile structure \cite{khitrova1968, bolzan1995,wyckoff1963}.}
\label{ionic_bond_strength_5d}
\resizebox{\textwidth}{!}{%
\begin{tabular}{cccccccc}
\hline
Material                                                     & TaO$_{2}$                                                 & WO$_{2}$**                                                 & ReO$_{2}$**                                                 & OsO$_{2}$                                                   & IrO$_{2}$                                                   & PtO$_{2}$**                                                   & AuO$_{2}$*                                                  \\ \hline
Löwdin charges ($e$)                                         & \begin{tabular}[c]{@{}c@{}}Ta: 1.52\\ O: $-0.76$\end{tabular} & \begin{tabular}[c]{@{}c@{}}W: 1.56\\ O: $-0.78$\end{tabular} & \begin{tabular}[c]{@{}c@{}}Re: 1.48\\ O: $-0.74$\end{tabular} & \begin{tabular}[c]{@{}c@{}}Os: 1.44\\ O: $-0.72$\end{tabular} & \begin{tabular}[c]{@{}c@{}}Ir: 1.50\\ O: $-0.75$\end{tabular} & \begin{tabular}[c]{@{}c@{}}Pt: 1.54\\ O: $-0.77$\end{tabular} & \begin{tabular}[c]{@{}c@{}}Au: 1.64\\ O: $-0.82$\end{tabular} \\
$\chi_\mathrm{bond}$ & 0.38                                                        & 0.39                                                       & 0.37                                                        & 0.36                                                        & 0.37                                                        & 0.38                                                        & 0.41                                                        \\
$-E_\mathrm{M}$                                            & 3.22                                                        & 3.48                                                       & 3.06                                                        & 3.00                                                        & 3.23                                                        & 3.34                                                        & 3.78                                                        \\ \hline
\end{tabular}%
}
\end{table}

Based on the analysis of the $-$iCOHP values, discussed in the previous section, almost identical strength of the covalent M--O bonds was found for IrO$_2$ and RuO$_2$. The covalent bonding conditions in the bulk oxides can therefore not explain the different \textit{operando} stability properties of these catalysts under OER conditions. To complete the analysis of the bonding characteristics in the bulk oxides, the ionic lattice energy of the investigated rutile metal oxides was examined. Based on the effective lumped charges of the individual atom species obtained from the Löwdin partitioning of the DFT electron density, it was first assessed how well the formal oxidation states (M: +4 and O: $-2$) were met in the simulated MO$_{2}$ systems. In addition, the Madelung energy per M--O bond ($E_\mathrm{M}$) was calculated as an energetic measure of the ionic lattice energy, as described in the Methods section. Tables \ref{ionic_bond_strength_4d} and \ref{ionic_bond_strength_5d} list the Löwdin charges, the ratio $\chi_\mathrm{bond}$ as a descriptor for the degree of ionicity, and $E_\mathrm{M}$ for the 4d and 5d MO$_2$ series, respectively.\\

\begin{figure}[!t]
  \centering
  \includegraphics[width=1\linewidth]{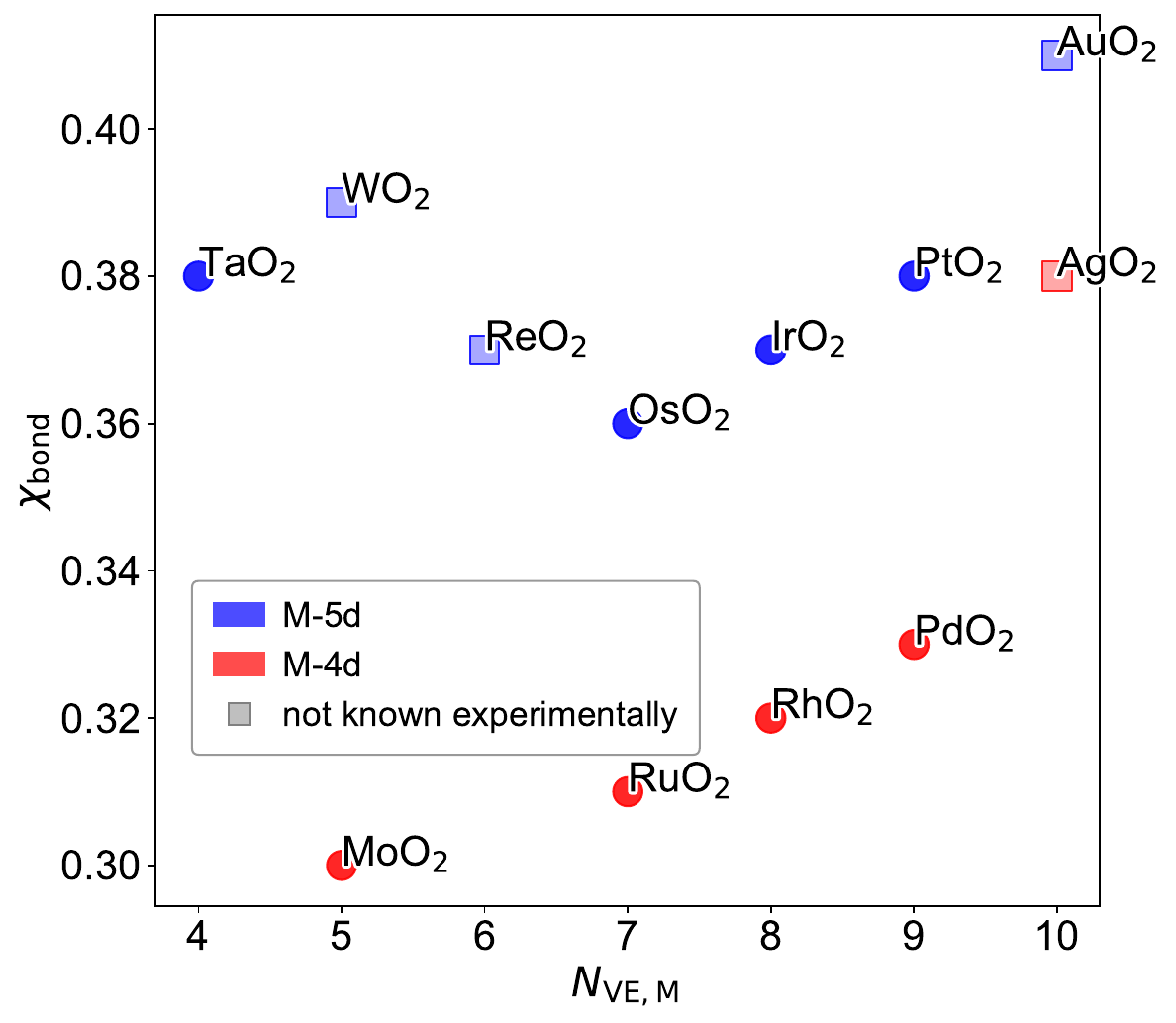}
\caption {Ionicity as $\chi$) in relation to the  $N_{\mathrm{VE,M}}$ count. Dashed circles around data points indicate artificial materials for which no experimental structural data is available.}
\label{u_ionicity_valence_electrons}
\end{figure}

\begin{figure}[!t]
  \centering
  \includegraphics[width=1\linewidth]{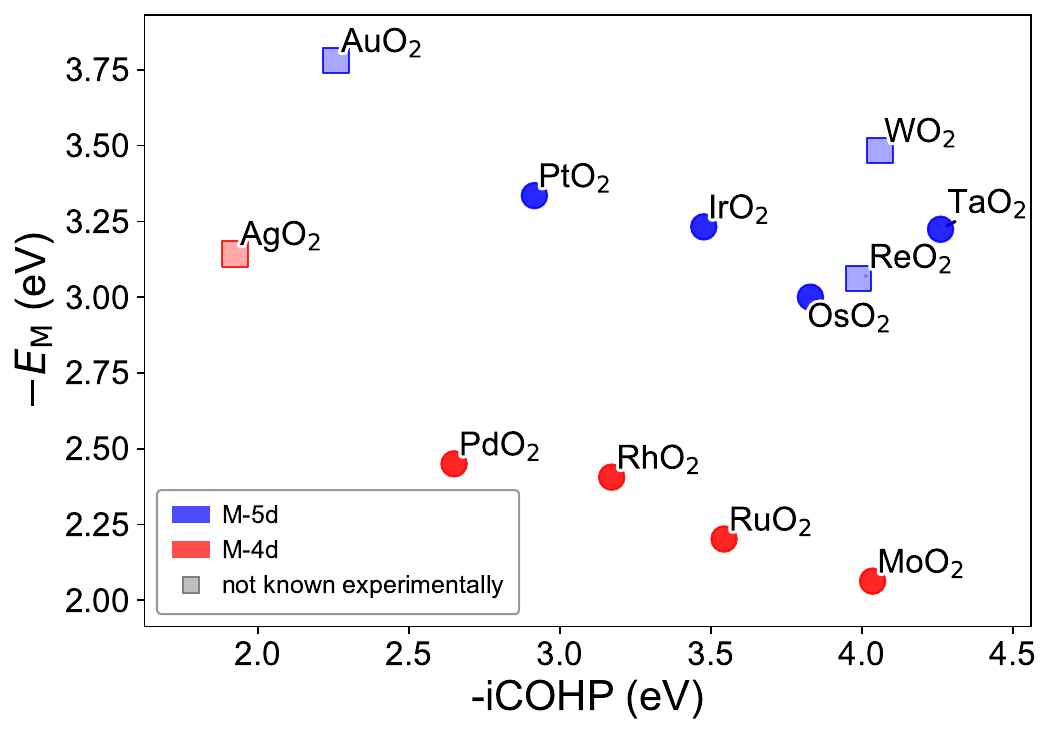}
\caption{The energetic descriptors for the ionic ($-E_\mathrm{M}$) and covalent ($-$iCOHP) bond strength plotted against each other. The data points are color-coded according to the d-orbital series of the respective metal (4d: red, 5d: blue). Dashed circles indicate artificial materials for which no experimental structural data is available.}
\label{u_icohp}
\end{figure}

Figure \ref{u_ionicity_valence_electrons} shows the relationship between $N_{\mathrm{VE,M}}$ and $\chi_\mathrm{bond}$. It can further be observed that the 5d metal oxides display higher ionicity than the 4d oxides. Figure \ref{u_icohp} reveals that all investigated 5d MO$_{2}$ oxides exhibit high ionic lattice energies, whereas the 4d oxides, with the exception of the artificial AgO$_{2}$, display comparatively lower ionic lattice energies. An inverse correlation between covalent bond strength and ionic lattice energy is observed within each series. This trend reflects the degree of electron localization. It underscores the importance of considering both the atomic and molecular orbital perspectives when assessing the total M--O bond strength, as M--O bond possess both covalent and ionic character. A M--O bond may be considered strong only when both, covalent and ionic, contributions are significant. Comparing IrO$_{2}$ and RuO$_{2}$ in Figure \ref{u_icohp}, IrO$_{2}$ exhibits significantly stronger ionic lattice energy than RuO$_{2}$, which may account for the generally stronger M--O bonds in IrO$_{2}$ suggested in previous studies \cite{binninger2022,tang2025,galyamin2023}. 

\section{Discussion}

This study examined the extent to which covalent bonding and ionic lattice energy contribute to the M--O bond in rutile oxides, and to assess whether the stability difference between IrO$_{2}$ and RuO$_{2}$ observed under OER \textit{operando} conditions \cite{cherevko2016,escaleralopez2021} can be related to relative strenghts of the M--O bonds in the bulk oxides.\\

The present study arrives at a somewhat different picture than the comparable analysis conducted by Tang et al. \cite{tang2025}, who concluded that IrO$_{2}$ is characterized by covalent bonding with substantial electron filling in the Ir--O bonds, while RuO$_{2}$ is dominated by ionic lattice energy with significant electron delocalization. The highly covalent bonding character of IrO$_{2}$ is confirmed in the present work, as evidenced in Figure \ref{icohp_valence_electrons}. However, RuO$_{2}$ is found to exhibit an essentially equivalent degree of covalent M--O bonding. The similar covalent M--O bonding observed in IrO$_{2}$ and RuO$_{2}$ can be rationalized as follows: Although IrO$_{2}$ accommodates a greater number of $N_{\mathrm{VE,M}}$ in antibonding molecular orbitals, the spatial extension of the 5d orbitals of Ir promotes constructive overlap with the O p-orbitals. A constructive orbital overlap results in the formation of bonding molecular orbitals and thereby compensating for the higher antibonding occupancy. The second conclusion of Tang et al. regarding the ionic character of RuO$_{2}$ is not corroborated by the present findings. As shown in Figures \ref{u_ionicity_valence_electrons} and \ref{u_icohp}, the calculated $\chi_\mathrm{bond}$ and $E_\mathrm{M}$ values indicate that 4d metal oxides generally exhibit lower ionicity, which is reflected in their comparatively weaker ionic lattice energies relative to the 5d metal oxides. The pronounced difference in ionic lattice energy between 4d and 5d metal oxides suggests that ionicity may be a decisive factor when assessing total M--O bond stability. While the covalent character constitutes a substantial contribution to the overall M--O bond strength, it does not appear to be the determining factor for bonding conditions in bulk rutile oxides. A possible explanation for the difference in ionic lattice energy between 4d and 5d materials may lie in their different electron affinities. This may facilitate electron transfer from the metal to oxygen in 5d oxides and thereby contribute to the stronger ionic lattice energy.\\

Figure \ref{u_icohp} provides a basis for rationalizing the  bulk lattice stability of IrO$_{2}$: both the covalent and ionic contributions play a significant role in the Ir--O bond. Recent computational studies tried to capture the total M--O bond strength through assessing the d-band center of the metal oxide of interest \cite{tang2025,maraschin2026}. The present study offers a complementary perspective on covalent bond strength and, by extension, on total M--O bond strength, through the lens of molecular orbital theory. As shown in Figure \ref{pdos_cohp} and \ref{icohp_valence_electrons}, an antibonding molecular orbital is located at $E_\mathrm{F}$. This is in line to the observations made in a previous study, where an antibonding $\pi^{*}$ molecular orbital was observed near E$_\mathrm{F}$ \cite{ping2015}.\\ 

The insights gained from the bulk electronic structure of the 4d and 5d metal oxides can be extended to surface and interface stability under OER \textit{operando} conditions. Strong M--O bonds in OER catalyst reduce the probability of M--O bond breaking at the surface, which in turn contributes to a higher overall stability of the catalyst. This is consistent with a previous study on the IrO$_{2}$ surface, where the binuclear mechanism involving the Ir--OOOO--Ir transition state was found to be energetically more favorable, as it does not involve any M--O bond cleavage \cite{binninger2022}. This hypothesis is further supported by the results presented in Figure \ref{u_icohp}, where IrO$_{2}$ is shown to exhibit both strong covalent and ionic contributions to the total M--O bond strength. Based on our studies OsO$_{2}$ may be a reasonable OER catalyst material, given that it yields results comparable to IrO$_{2}$ and is experimentally known to adopt the rutile structure \cite{goldschmidt1926}. Other experimental studies  have identified  OsO$_{2}$ as the most active rutile oxide due to its low overpotential. At the same time, OsO$_{2}$ was found to be unstable, and therefore not suitable as OER catalyst material \cite{danilovic2014} which is slightly contradictory to the high covalent and ionic bond strength suggested by our computational results. However, it should be noted that the present analysis is based on metal-oxygen bonding conditions, and thus intrinsic stability of the respective rutile phase only. The actual stability of a specific oxide under OER conditions equally depends on the thermodynamic competition with possible degradation products, e.g., dissolved transition metal complexes \cite{maraschin2026}. A complete computational assessment of operando stability will therefore require accounting for the product side and the kinetics of possible degradation pathways. On the other hand, our results also suggest TaO$_{2}$ as a candidate for high intrinsic stability. It is interesting to note that, indeed, tantalum was found to have exert a stabilizing effect in rutile mixed Ta-Ru-oxides \cite{zhang2025}.   

\section{Conclusion}

The presented computational study focused on comparing the strength of M--O bonds bulk IrO$_{2}$ and RuO$_{2}$ through the molecular orbital framework to explore the stability differences observed in OER catalysis. Findings of this study suggest that the stability difference between IrO$_{2}$ and RuO$_{2}$ may originate from differences in their total M--O bond strength. Importantly, this difference does not arise primarily from the covalent contribution, as commonly assumed \cite{tang2025}, but rather from the markedly strong ionic bonding in IrO$_{2}$ relative to RuO$_{2}$. Furthermore, a strong correlation was found between the covalent bond strength descriptor $-$iCOHP and the number of valence electrons of the metal $N_{\mathrm{VE,M}}$, across the selected set of 4d and 5d metal oxides. This correlation extends to the occupation of the antibonding $\pi^{*}$ molecular orbital: a maximum in $-$iCOHP is observed as $N_{\mathrm{VE,M}}=$ 4 to 5, beyond which progressive filling of the antibonding $\pi^{*}$ orbital leads to a monotonic decrease in covalent bond strength.\\

In future work, the insights obtained from the bulk metal oxide systems will be applied to the IrO$_{2}$(110) and RuO$_{2}$(110) surfaces, with the aim of assessing whether the surface M--O bonds exhibit analogous or distinctly different behavior compared to their bulk counterparts. In addition, the molecular orbital framework developed in this study will be leveraged to guide the design of Ir-based mixed metal oxides, targeting materials with reduced antibonding orbital occupancy and high ionic lattice energy, with the goal of identifying catalysts with total M--O bond strengths comparable to or exceeding that of IrO$_{2}$.

\section{Acknowledgements}

The authors acknowledge the financial support from the Federal Ministry of Science and Education (BMBF) under the German–Canadian Materials Acceleration Centre (GC-MAC) grant number 01DM21001A. The authors gratefully acknowledge the Gauss Centre for Supercomputing e.V. (www.gauss-centre.eu) for funding this project by providing computing time through the John von Neumann Institute for Computing (NIC) on the GCS Supercomputer JUWELS at Jülich Supercomputing Centre (JSC). The authors gratefully acknowledge Dr. Peter Müller and Prof. Dr. Richard Dronskowski (Institute of Inorganic Chemistry, RWTH Aachen University) for valuable discussions regarding COHP analysis and the use of LOBSTER. The authors thank Dr. Dustin Vivod for constructive advice on manuscript preparation.

\bibliography{ref}

\end{document}


\doublespacing
\section{pDOS and COHP for studied rutile oxides}

\begin{figure}[H]
  \centering
  \includegraphics[width=0.9\linewidth]{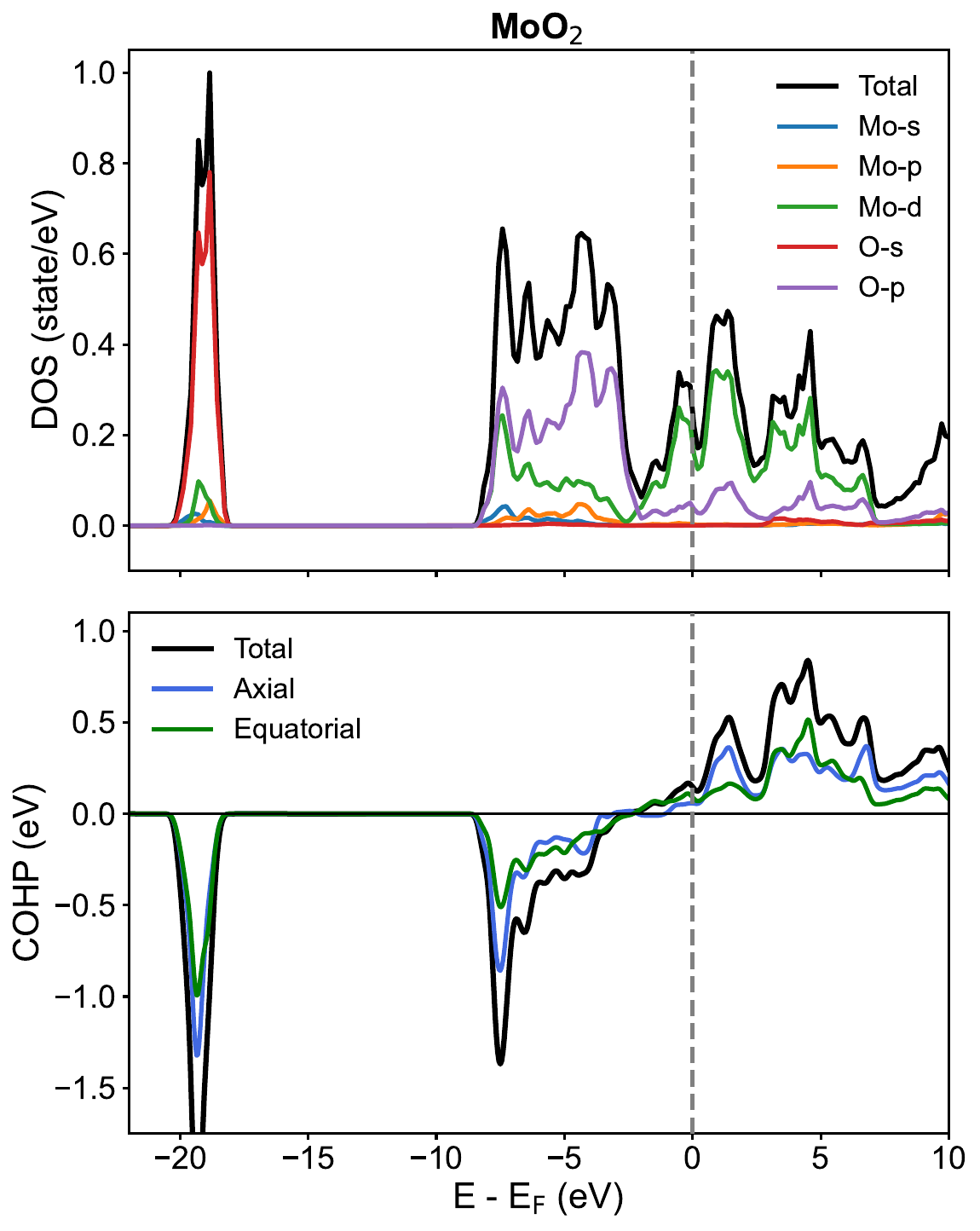}
\caption{Top: Total DOS and pDOS for rutile MoO$_{2}$ in the vicinity of $E_\mathrm{F}$ (dashed-line). The pDOS resolves contributions from metal s, p, d orbitals and oxygen s, p orbital, presented over a common energy window for direct comparison across both materials. Bottom: Corresponding COHP curves for the same energy range, showing the total M--O contribution together with separate traces for axial and equatorial M--O bonds in each material.}
\label{pdos_cohp_mo}
\end{figure}

\begin{figure}[H]
  \centering
  \includegraphics[width=0.9\linewidth]{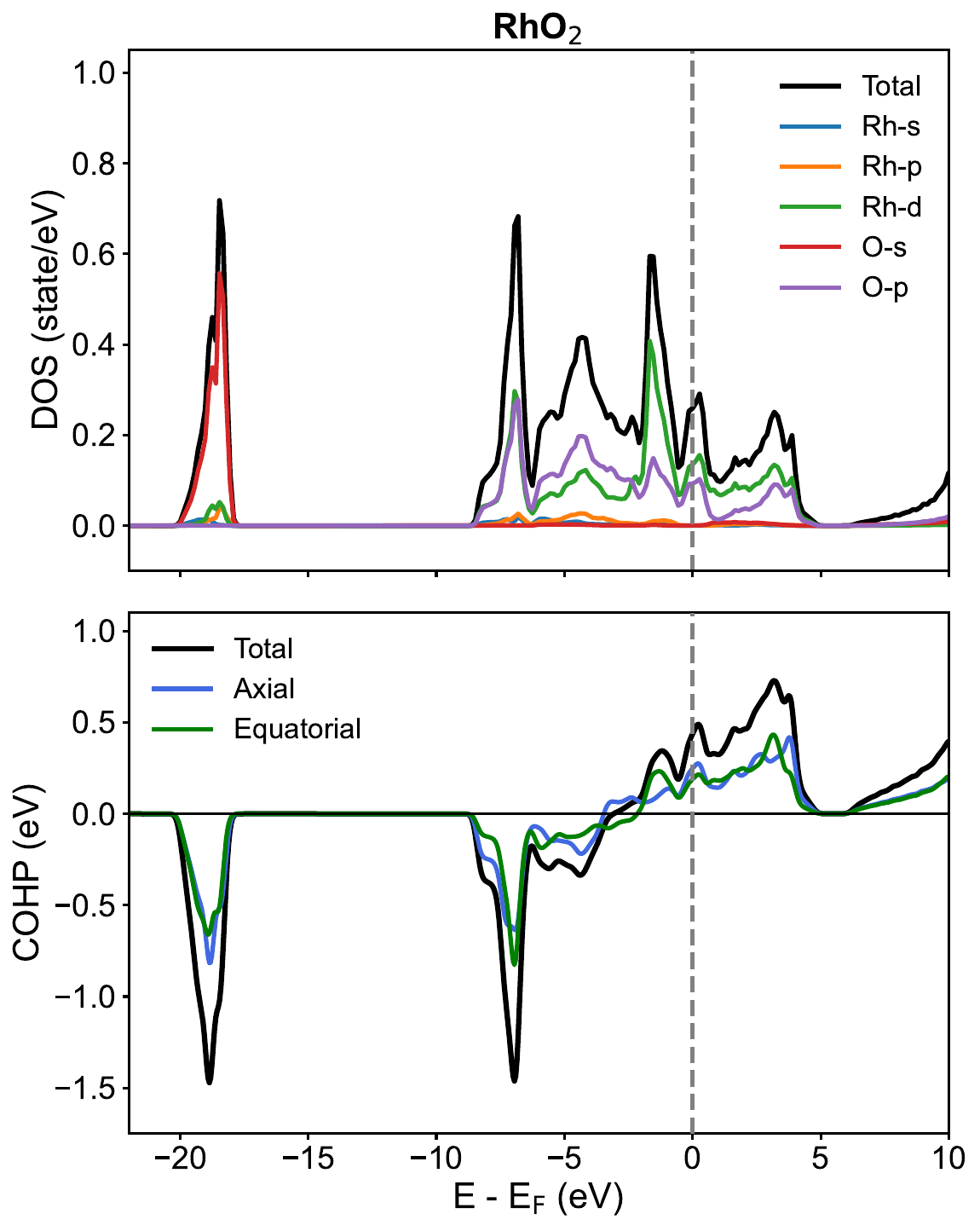}
\caption{Top: Total DOS and pDOS for rutile RhO$_{2}$ in the vicinity of $E_\mathrm{F}$ (dashed-line). The pDOS resolves contributions from metal s, p, d orbitals and oxygen s, p orbital, presented over a common energy window for direct comparison across both materials. Bottom: Corresponding COHP curves for the same energy range, showing the total M--O contribution together with separate traces for axial and equatorial M--O bonds in each material.}
\label{pdos_cohp_rh}
\end{figure}

\begin{figure}[H]
  \centering
  \includegraphics[width=0.9\linewidth]{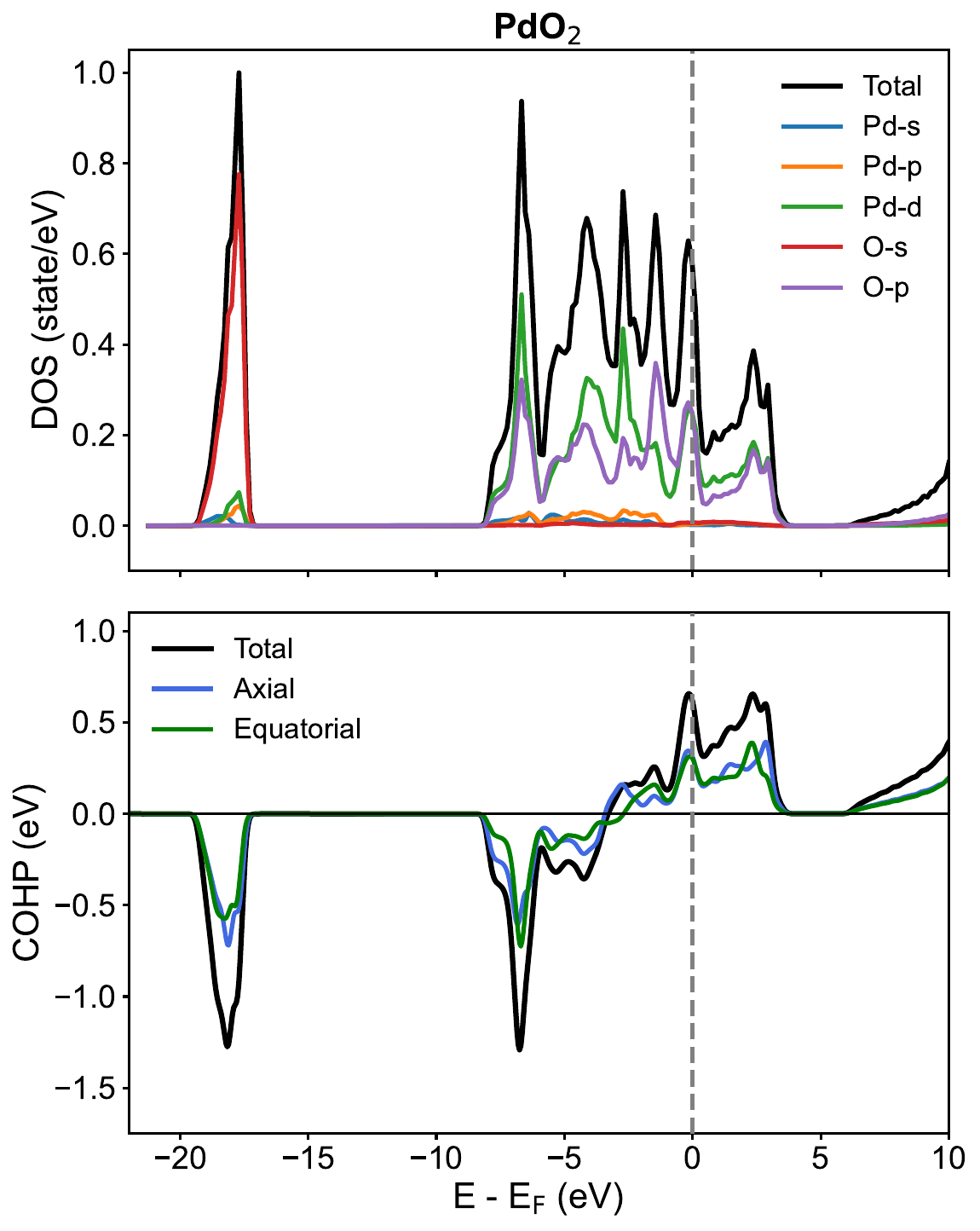}
\caption{Top: Total DOS and pDOS for rutile PdO$_{2}$ in the vicinity of $E_\mathrm{F}$ (dashed-line). The pDOS resolves contributions from metal s, p, d orbitals and oxygen s, p orbital, presented over a common energy window for direct comparison across both materials. Bottom: Corresponding COHP curves for the same energy range, showing the total M--O contribution together with separate traces for axial and equatorial M--O bonds in each material.}
\label{pdos_cohp_pd}
\end{figure}

\begin{figure}[H]
  \centering
  \includegraphics[width=0.9\linewidth]{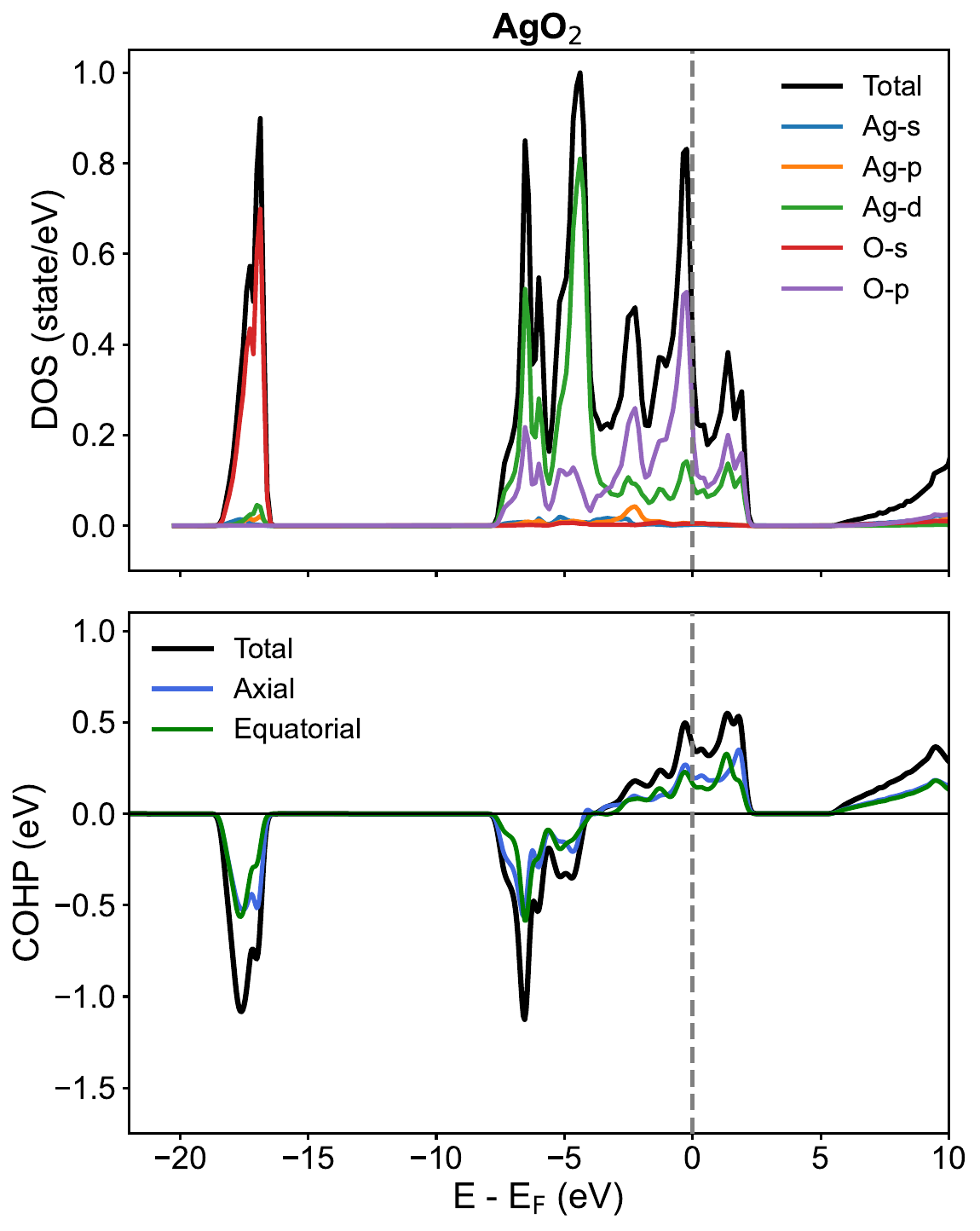}
\caption{Top: Total DOS and pDOS for rutile AgO$_{2}$ in the vicinity of $E_\mathrm{F}$ (dashed-line). The pDOS resolves contributions from metal s, p, d orbitals and oxygen s, p orbital, presented over a common energy window for direct comparison across both materials. Bottom: Corresponding COHP curves for the same energy range, showing the total M--O contribution together with separate traces for axial and equatorial M--O bonds in each material.}
\label{pdos_cohp_ag}
\end{figure}

\begin{figure}[H]
  \centering
  \includegraphics[width=0.9\linewidth]{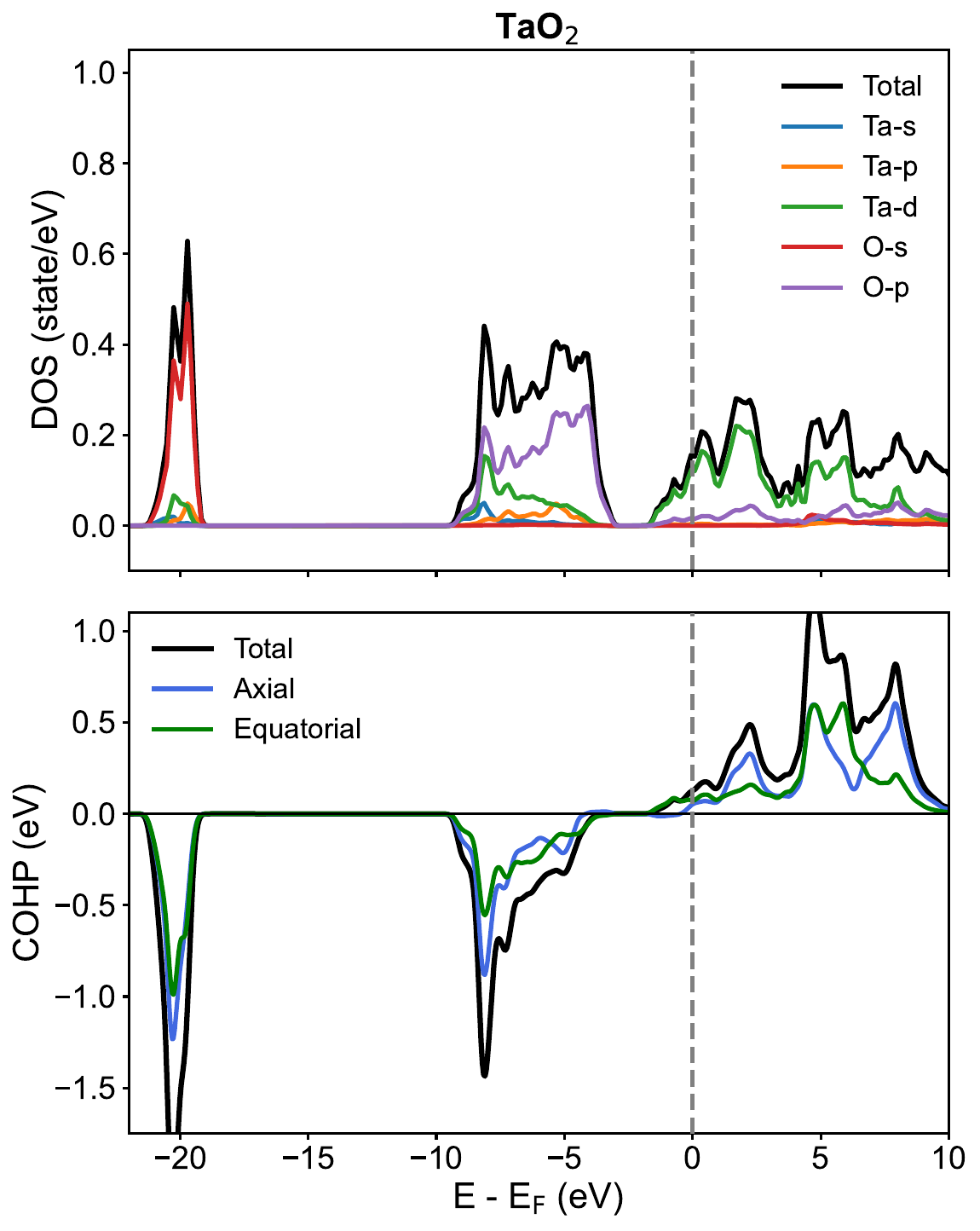}
\caption{Top: Total DOS and pDOS for rutile TaO$_{2}$ in the vicinity of $E_\mathrm{F}$ (dashed-line). The pDOS resolves contributions from metal s, p, d orbitals and oxygen s, p orbital, presented over a common energy window for direct comparison across both materials. Bottom: Corresponding COHP curves for the same energy range, showing the total M--O contribution together with separate traces for axial and equatorial M--O bonds in each material.}
\label{pdos_cohp_ta}
\end{figure}

\begin{figure}[H]
  \centering
  \includegraphics[width=0.9\linewidth]{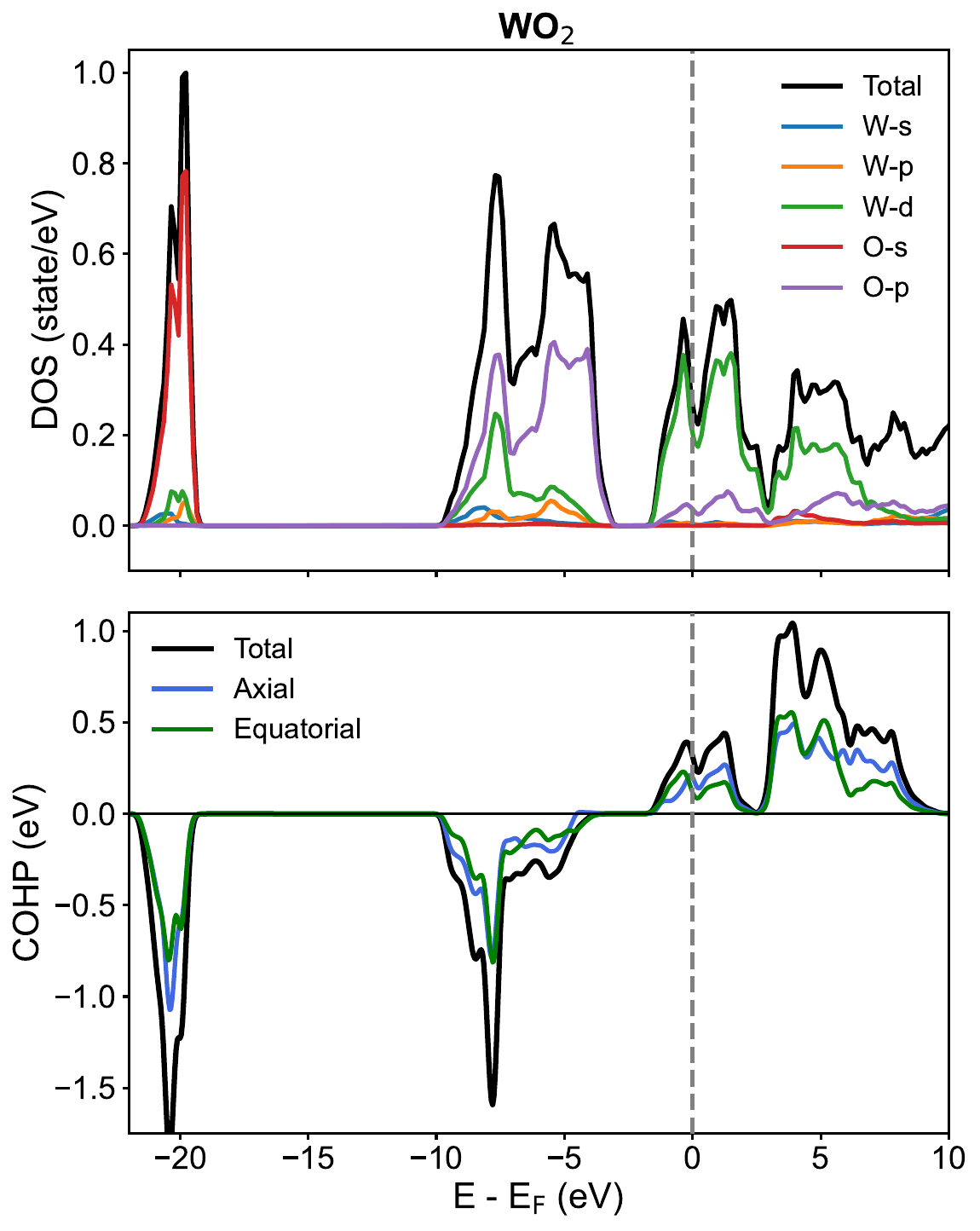}
\caption{Top: Total DOS and pDOS for rutile WO$_{2}$ in the vicinity of $E_\mathrm{F}$ (dashed-line). The pDOS resolves contributions from metal s, p, d orbitals and oxygen s, p orbital, presented over a common energy window for direct comparison across both materials. Bottom: Corresponding COHP curves for the same energy range, showing the total M--O contribution together with separate traces for axial and equatorial M--O bonds in each material.}
\label{pdos_cohp_w}
\end{figure}

\begin{figure}[H]
  \centering
  \includegraphics[width=0.9\linewidth]{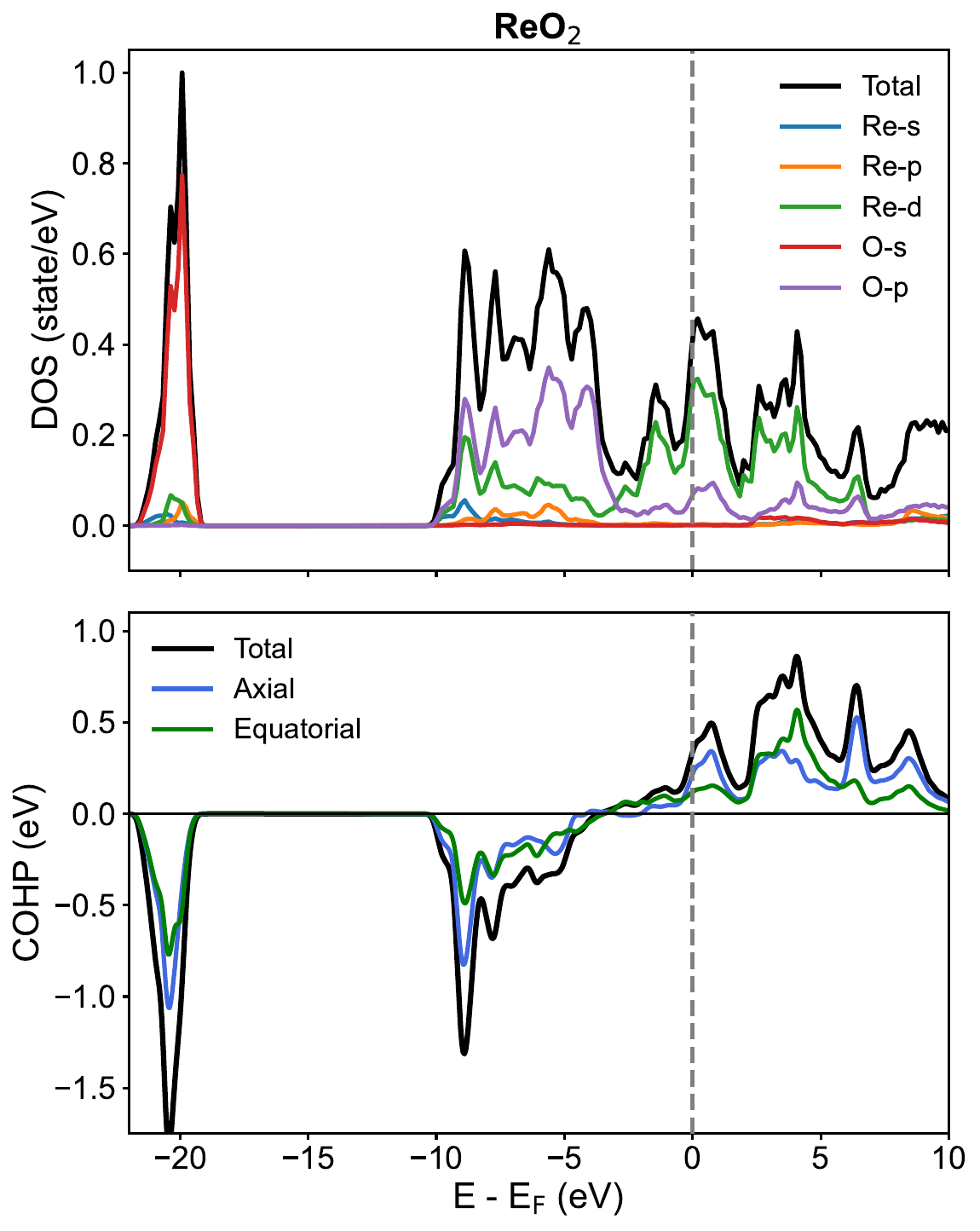}
\caption{Top: Total DOS and pDOS for rutile ReO$_{2}$ in the vicinity of $E_\mathrm{F}$ (dashed-line). The pDOS resolves contributions from metal s, p, d orbitals and oxygen s, p orbital, presented over a common energy window for direct comparison across both materials. Bottom: Corresponding COHP curves for the same energy range, showing the total M--O contribution together with separate traces for axial and equatorial M--O bonds in each material.}
\label{pdos_cohp_re}
\end{figure}

\begin{figure}[H]
  \centering
  \includegraphics[width=0.9\linewidth]{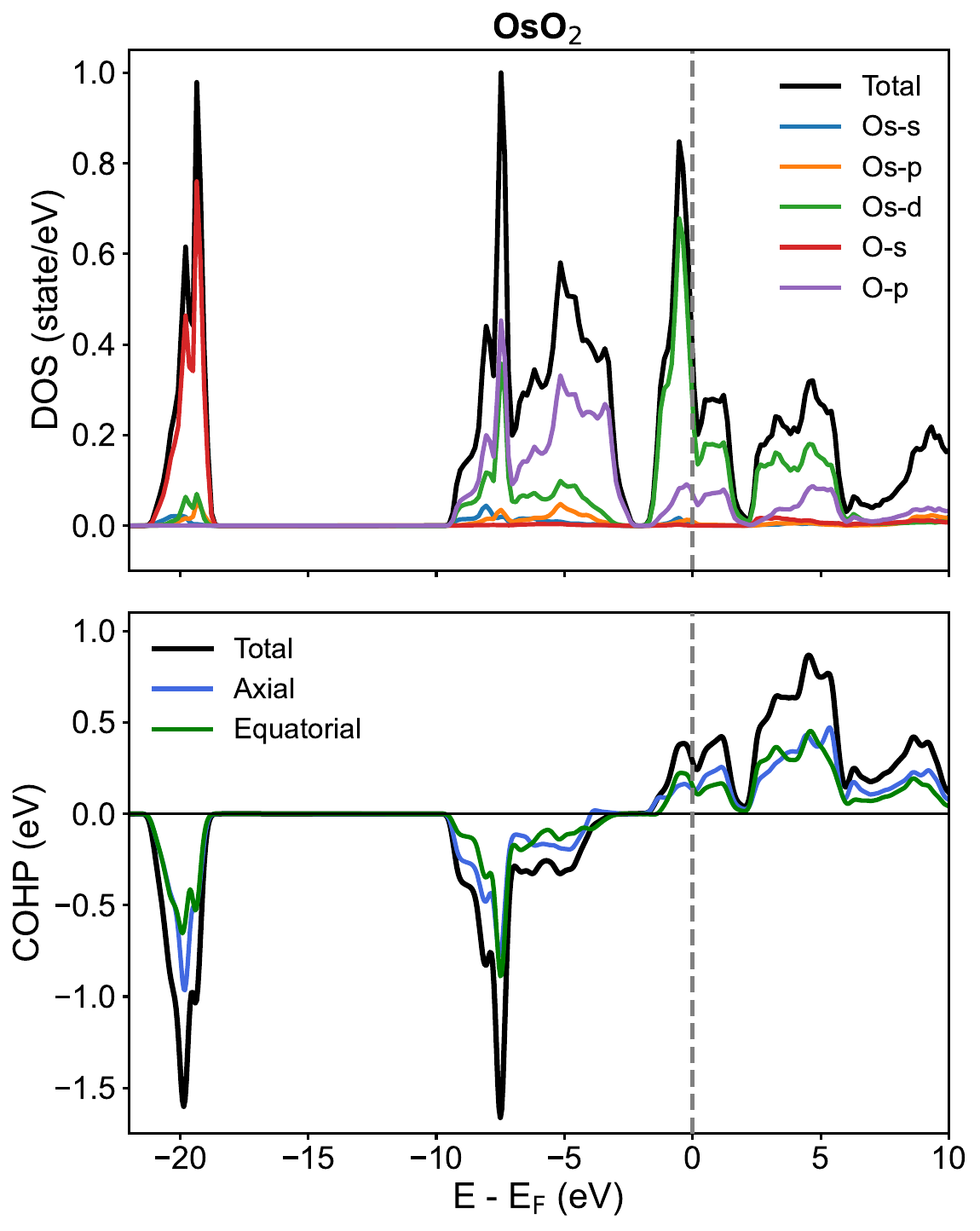}
\caption{Top: Total DOS and pDOS for rutile OsO$_{2}$ in the vicinity of $E_\mathrm{F}$ (dashed-line). The pDOS resolves contributions from metal s, p, d orbitals and oxygen s, p orbital, presented over a common energy window for direct comparison across both materials. Bottom: Corresponding COHP curves for the same energy range, showing the total M--O contribution together with separate traces for axial and equatorial M--O bonds in each material.}
\label{pdos_cohp_os}
\end{figure}

\begin{figure}[H]
  \centering
  \includegraphics[width=0.9\linewidth]{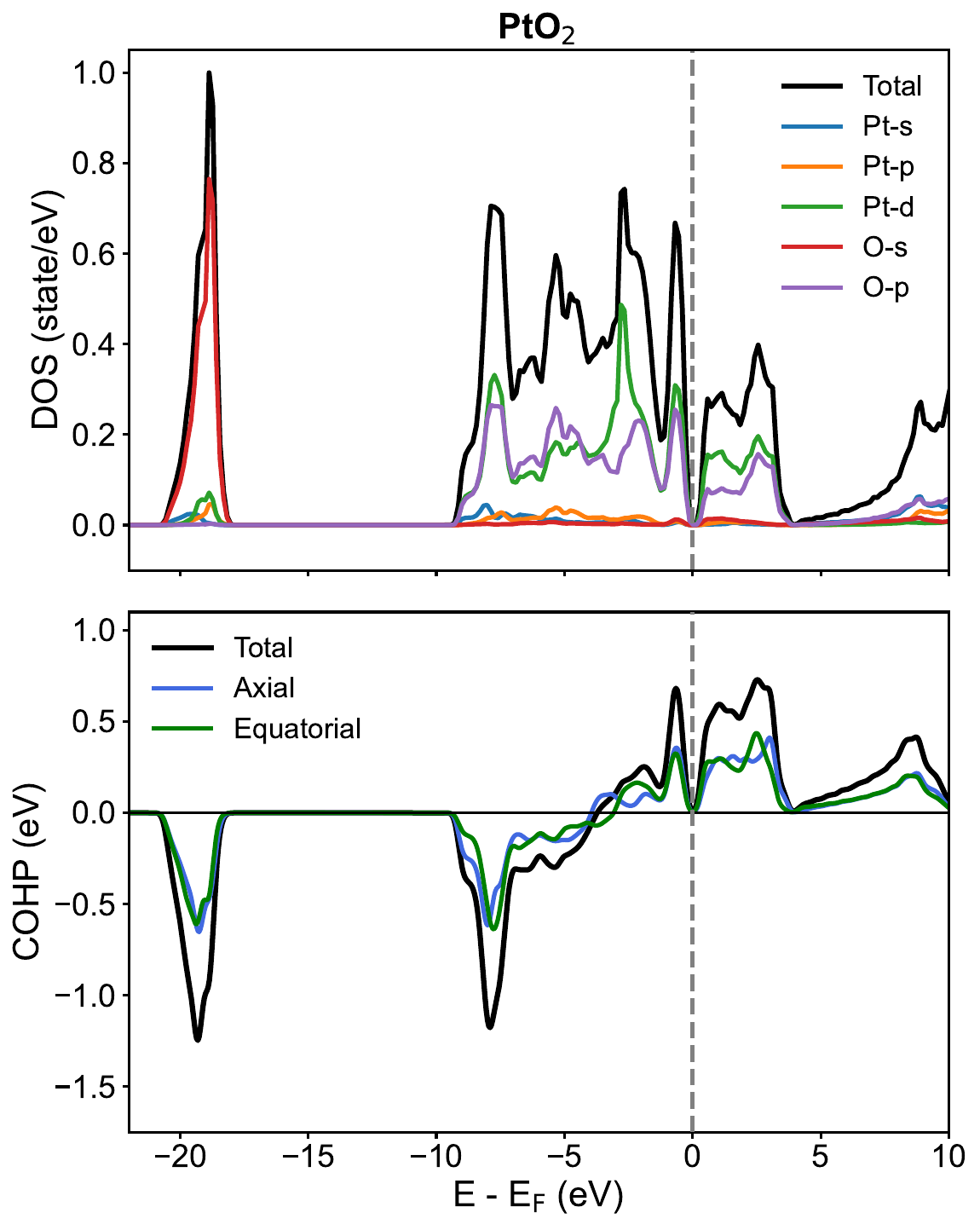}
\caption{Top: Total DOS and pDOS for rutile PtO$_{2}$ in the vicinity of $E_\mathrm{F}$ (dashed-line). The pDOS resolves contributions from metal s, p, d orbitals and oxygen s, p orbital, presented over a common energy window for direct comparison across both materials. Bottom: Corresponding COHP curves for the same energy range, showing the total M--O contribution together with separate traces for axial and equatorial M--O bonds in each material.}
\label{pdos_cohp_pt}
\end{figure}

\begin{figure}[H]
  \centering
  \includegraphics[width=0.9\linewidth]{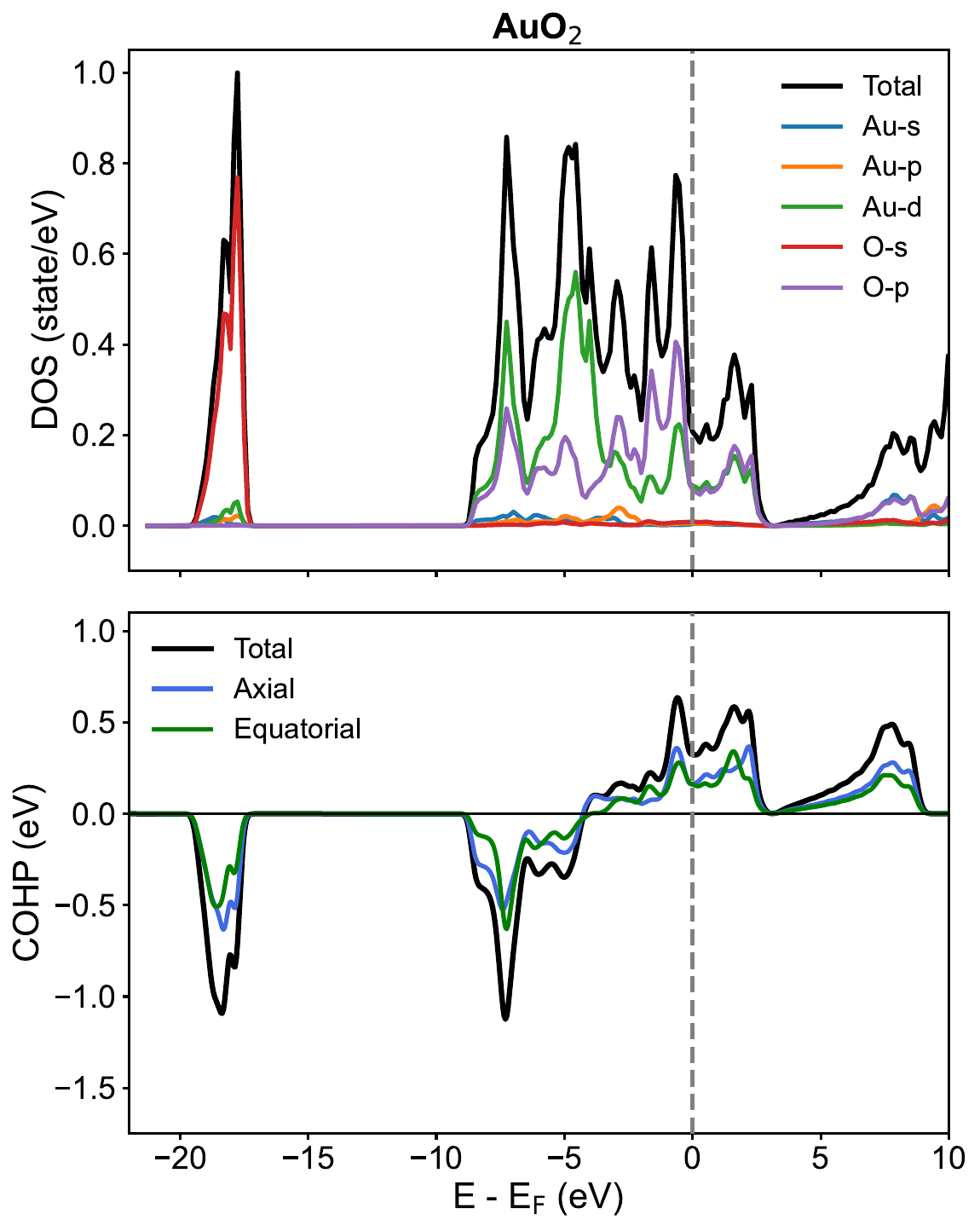}
\caption{Top: Total DOS and pDOS for rutile AuO$_{2}$ in the vicinity of $E_\mathrm{F}$ (dashed-line). The pDOS resolves contributions from metal s, p, d orbitals and oxygen s, p orbital, presented over a common energy window for direct comparison across both materials. Bottom: Corresponding COHP curves for the same energy range, showing the total M--O contribution together with separate traces for axial and equatorial M--O bonds in each material.}
\label{pdos_cohp_au}
\end{figure}

Figures \ref{pdos_cohp_mo} to \ref{pdos_cohp_au} show the pDOS and COHP of the studied rutile oxides from the 4d and 5d series. In general, all materials exhibit a similar overall shape of the pDOS and COHP curves, which makes the electronic structure comparable across the series. One notable difference lies in the magnitude of COHP contributions: for some materials, the COHP value is weak even where atomic orbital contributions are present in a given energy range, indicating that these states contribute little to the covalent bond strength (Figure \ref{pdos_cohp_ag}, Figure \ref{pdos_cohp_pt}, Figure \ref{pdos_cohp_au}). A further difference is the energy range of the occupied molecular orbitals, which was already discussed in the context of IrO$_{2}$ and RuO$_{2}$.

\section{d-band center and atomic orbital overlap}

\begin{figure}[H]
  \centering
  \includegraphics[width=1\linewidth]{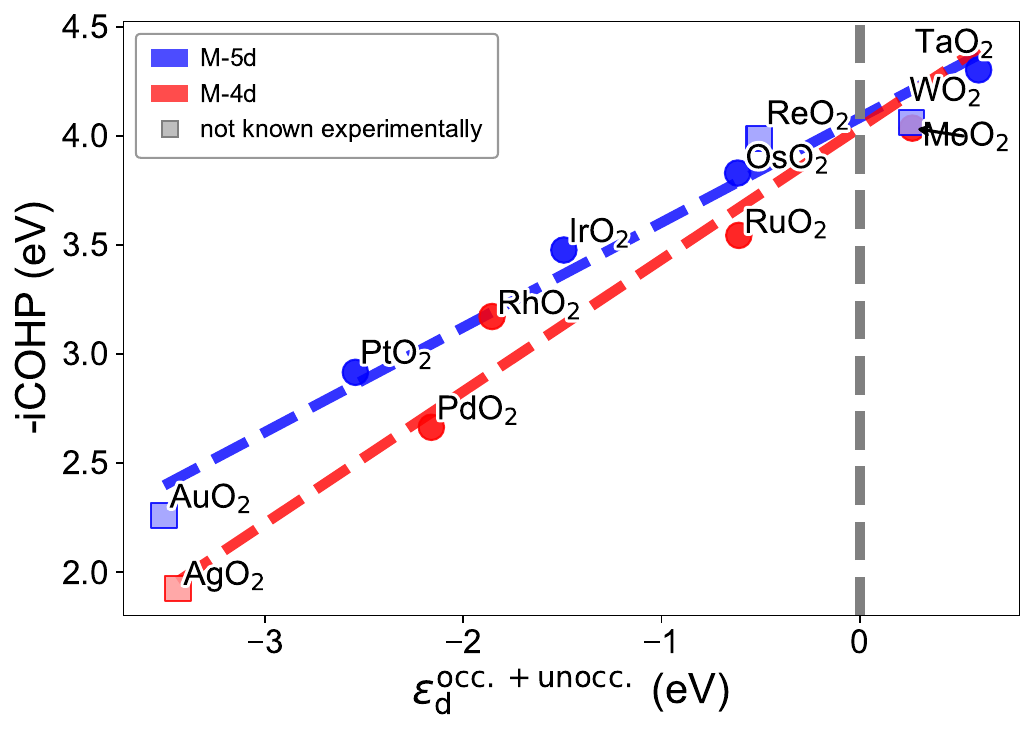}
\caption{$-$iCOHP versus $\varepsilon_\mathrm{d}^\mathrm{occ.+unocc.}$ for the rutile MO$_{2}$ materials. Data for 5d-metals are shown in blue, and data for 4d-metals are shown in red. $E_\mathrm{F}$ is shown at $E = 0\, \mathrm{eV}$. The dashed dots indicate the materials where no experimental data was available.}
\label{icohp_d_band_center}
\end{figure}

\begin{figure}[H]
  \centering
  \includegraphics[width=1\linewidth]{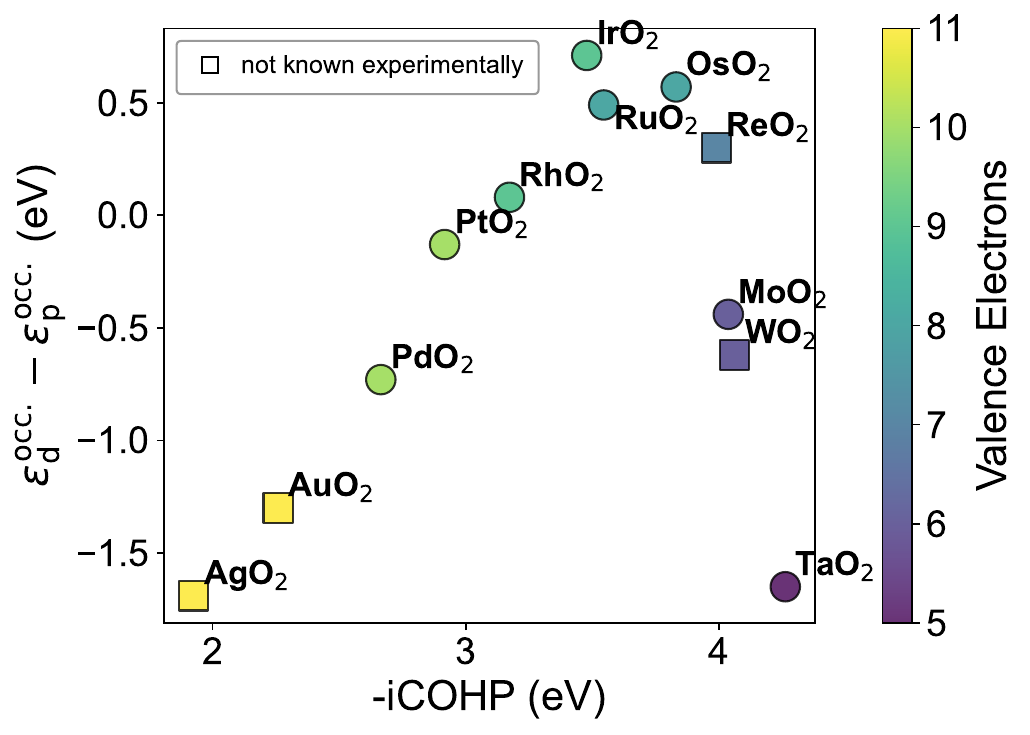}
\caption{$\varepsilon_\mathrm{d,M}^\mathrm{occ.} - \varepsilon_\mathrm{p,O}^\mathrm{occ.}$) calculated up to E$_\mathrm{F}$ over the covalent bond strength descriptor $-$iCOHP for the rutile MO$_{2}$ series. Data points are color-coded by the number of $N_{\mathrm{VE,M}}$ in the metal d-orbital. Dashed circles around a data point indicate that no experimental structural data is available for the respective material.}
\label{covalency_ionicity}
\end{figure}

 Figure \ref{icohp_d_band_center} and Figure \ref{covalency_ionicity} show the assessment of the covalent bond strength from an atomic orbital perspective. Figure \ref{pdos_cohp} in the main text indicate that the d-orbitals of the metal shift in energy depending on the number of $N_{\mathrm{VE,M}}$ in the d-orbital. To further examine this trend, the d-band center $\varepsilon_\mathrm{d}^\mathrm{occ.+unocc.}$ was calculated for all materials in order to assess its relationship to $-$iCOHP and its suitability as a bulk descriptor. $\varepsilon_\mathrm{d}^\mathrm{occ.+unocc.}$ as it is well established as a surface descriptor for electrocatalytic materials \cite{lima2007,hammer2000,ruban1997,norskov2009,norskov2011,vojvodic2014}.\\

Figure \ref{icohp_d_band_center} plots $-$iCOHP versus $\varepsilon_\mathrm{d}^\mathrm{occ.+unocc.}$ for the different oxides. For WO$_{2}$, which exhibits magnetic behavior and distinct spin-up and spin-down DOS, the spin-averaged $\varepsilon_\mathrm{d}^\mathrm{occ.+unocc.}$ is used. Both 4d and 5d series display linear correlations: lower (more negative $\varepsilon_\mathrm{d}^\mathrm{occ.+unocc.}$) values correspond to smaller $-$iCOHP magnitudes and thus weaker covalent bonds. The 5d fit yields:
\begin{equation}
    -\mathrm{iCOHP} = 0.481\varepsilon_\mathrm{d}^\mathrm{occ.+unocc.} + 4.084\,\mathrm{eV},
\end{equation}

\noindent while the 4d fit gives:

\begin{equation}
    -\mathrm{iCOHP} = 0.606\varepsilon_\mathrm{d}^\mathrm{occ.+unocc.} + 4.039\,\mathrm{eV}.
\end{equation}

Among 5d oxides, AuO$_{2}$ exhibits weakest covalent bonds ($\varepsilon_\mathrm{d}^\mathrm{occ.+unocc.} = -3.51\,\mathrm{eV}$) and TaO$_{2}$ the highest ($\varepsilon_\mathrm{d}^\mathrm{occ.+unocc.} = 0.60\,\mathrm{eV}$). For 4d oxides, AgO$_{2}$ shows the lowest ($\varepsilon_\mathrm{d}^\mathrm{occ.+unocc.} = -3.44\,\mathrm{eV}$) and MoO$_{2}$ the strongest ($\varepsilon_\mathrm{d}^\mathrm{occ.+unocc.} = 0.27\,\mathrm{eV}$) covalent bonds. When comparing IrO$_{2}$ and RuO$_{2}$, $\varepsilon_\mathrm{d}^\mathrm{occ.+unocc.}$ of RuO$_{2}$ ($\varepsilon_\mathrm{d} = -0.61\,\mathrm{eV}$) lies higher in energy than that of IrO$_{2}$ ($\varepsilon_\mathrm{d} = -1.492\,\mathrm{eV}$). However, because Ir belongs to the 5d series and Ru to the 4d series, no large difference in $-$iCOHP is observed, indicating that the M--O covalent bond strength is similar in both materials despite their different d-band center positions.\\

Figure \ref{icohp_d_band_center} indicates that a relationship exists between the metal d-band center and $-$iCOHP. To further examine this, $\varepsilon_\mathrm{d,M}^\mathrm{occ.} - \varepsilon_\mathrm{p,O}^\mathrm{occ.}$ was plotted as a function of $-$iCOHP, as shown in Figure \ref{covalency_ionicity}. The overall distribution exhibits a volcano-like shape: starting from the lowest $-$iCOHP values, $\varepsilon_\mathrm{d,M}^\mathrm{occ.} - \varepsilon_\mathrm{p,O}^\mathrm{occ.}$ increases approximately linearly up to $\approx 3.5\,$eV, after which it decreases linearly with a steeper slope. Comparing the $N_{\mathrm{VE,M}}$ count with the orbital overlap and covalent bond strength reveals a peak at approximately eight to nine $N_{\mathrm{VE,M}}$, where both strong covalent bonds (large $-$iCOHP) and substantial orbital overlap ($\varepsilon_\mathrm{d,M}^\mathrm{occ.} - \varepsilon_\mathrm{p,O}^\mathrm{occ.}$ close to zero) coincide. In contrast, the values of $\varepsilon_\mathrm{d,M}^\mathrm{occ.} - \varepsilon_\mathrm{p,O}^\mathrm{occ.}$ do not show a clear relationship to the covalent bond strength. The most negative values of $\varepsilon_\mathrm{d,M}^\mathrm{occ.} - \varepsilon_\mathrm{p,O}^\mathrm{occ.}$ are observed for the materials with either the lowest or highest $N_{\mathrm{VE,M}}$ count, which suggests that orbital overlap plays a less significant role in determining the covalent bond strength.\\

AgO$_{2}$ ($\varepsilon_\mathrm{d,M}^\mathrm{occ.} - \varepsilon_\mathrm{p,O}^\mathrm{occ.}$ = -1.69\,eV) and TaO$_{2}$ ($\varepsilon_\mathrm{d,M}^\mathrm{occ.} - \varepsilon_\mathrm{p,O}^\mathrm{occ.}$ = -1.65\,eV) display the lowest orbital overlap, yet differ in covalent bond strength: AgO$_{2}$ shows weak covalent bonds ($-$iCOHP = 1.93\,eV), whereas TaO$_{2}$ exhibits strong covalent bonds ($-$iCOHP = 4.30\,eV). Materials combining strong covalent bonds with substantial orbital overlap include OsO$_{2}$ ($\varepsilon_\mathrm{d,M}^\mathrm{occ.} - \varepsilon_\mathrm{p,O}^\mathrm{occ.}$ = 0.57\,eV, $-$iCOHP = 3.83\,eV), RuO$_{2}$ ($\varepsilon_\mathrm{d,M}^\mathrm{occ.} - \varepsilon_\mathrm{p,O}^\mathrm{occ.}$ = 0.49\,eV, $-$iCOHP = 3.54\,eV), and IrO$_{2}$ ($\varepsilon_\mathrm{d,M}^\mathrm{occ.} - \varepsilon_\mathrm{p,O}^\mathrm{occ.}$ = 0.71\,eV, $-$iCOHP = 3.48\,eV).\\

\bibliography{ref}